%% file: finalv3.tex
\documentclass[aps,amsmath,amssymb]{revtex4}
\usepackage{amsmath,amssymb}
\usepackage{graphicx}
\usepackage{dcolumn}
\usepackage{bm}
\usepackage{tabularx}
\usepackage{wick}

\newcommand{\bfm}[1]{\mbox{\boldmath$#1$}}
\newcommand{\Slash}[1]{{\ooalign{\hfil/\hfil\crcr$#1$}}}

  \def\Su(#1){{S^{u}_{#1}}}
  \def\Sd(#1){{S^{d}_{#1}}}
  \def\Ss(#1){{S^{s}_{#1}}}
  \def\S(#1){{S_{#1}}}

  \def\g{{\gamma}} 
  \def\q{{\overline q}}

  \def\h{{\overline h}}

   \def\Ci{C^{-1}}

  \def\gt_#1{(-C\g_{#1}\Ci)}
  \def\bra{\langle}
  \def\ket{\rangle}
  \def\qcon{\bra\q q\ket}

  \def\gc{\bra \frac{\alpha_s}{\pi}G^2\ket}

  \def\gcn{\bra G^2\ket}  
  \def\mixqn{\bra \q g G\cdot \sigma q\ket}

  \def\eps{\epsilon}
  \def\psla{p{\raise1pt\hbox{$\!\!/$}}}
  \def\qsla{q{\raise1pt\hbox{$\!\!\!/$}}}
  \def\knd(#1,#2){\delta_{#1#2}}

   \def\sig{\sigma}
   \def\Tr{{\rm Tr}}

\def\pp{P_+}

\def\fig#1{Fig.\ref{#1}}

\def\gev{{\rm GeV}}
\input{def}
\begin{document}
\preprint{}
\title{QCD sum rules
for quark-gluon three-body components in the $B$ meson
}

\author{Tetsuo NISHIKAWA}
\email{nishikawa@ryotokuji-u.ac.jp} 
\affiliation{%
Faculty of Health Science, Ryotokuji University, Urayasu, Chiba 279-8567, Japan
}
\author{Kazuhiro TANAKA}
\email{kztanaka@juntendo.ac.jp} 
\affiliation{%
Department of Physics, Juntendo University, Inzai, Chiba 270-1695, Japan\\
and
J-PARC Branch, KEK Theory Center,
Institute of Particle and Nuclear Studies,
High Energy Accelerator Research Organization (KEK),
203-1, Shirakata, Tokai, Ibaraki, 319-1106, Japan
}

\date{\today}

\begin{abstract}
We discuss the QCD sum rule calculation of the
heavy-quark effective theory parameters, $\lambda_E$ and $\lambda_H$,
which correspond to matrix elements representing quark-gluon three-body
components in the $B$-meson wavefunction.
We derive the sum rules for $\lambda_{E,H}$
calculating the new higher-order QCD corrections,
i.e., the order $\alpha_s$ radiative corrections
to the Wilson
coefficients associated with the dimension-5 quark-gluon mixed
condensates, and 
the
power corrections
due to the dimension-6 vacuum condensates.
We find that
the new radiative corrections 
significantly
improve the stability of the corresponding Borel sum rules
and lead to the reduction of the values of $\lambda_{E,H}$.
We also discuss the renormalization-group improvement for the sum rules
and present update 
on the values of $\lambda_{E,H}$.
\end{abstract}
\maketitle

\section{Introduction}
The $B$ mesons play distinguished roles
in exploring CP violation and the flavor sector
of the Standard Model.
In particular, the measurements of the $B$-meson decays 
can 
provide precise 
information on 
the relevant quark couplings~\cite{Antonelli:2009ws}.
Since the properties of those decays are also
influenced by 
the complicated strong-interaction effects
responsible for forming the $B$-meson,
as well as the final-state hadrons,
a better theoretical control 
of the nonperturbative effects inside the $B$ mesons is now becoming
very important.
This is also interesting in its own right
as understanding the properties of the simplest meson including a 
heavy quark.

In the heavy-quark limit based on 
$\Lambda_{\rm QCD}/m_B\ll 1$, $\Lambda_{\rm QCD}/m_b \ll 1$
with $m_B$ and $m_b$ being the masses of the $B$-meson and $b$-quark,
the matrix elements using a $B$-meson state
obey heavy-quark symmetry,
and are conveniently described by the heavy-quark effective theory (HQET)~\cite{neubert0}.
In this framework, fundamental properties of the $B$ mesons are
represented by the HQET parameters
that are defined as matrix elements of the 
relevant local operators, 
like the decay constant $F$~\cite{neubert0}:
\begin{equation}
\bra 0|\overline{q}\gamma_\rho \g_5h_v |\bar B(v)\ket
=iF(\mu)v_\rho .
\label{fmu}
\end{equation}
Here, $|\bar B(v)\ket$ is the $\bar B$-meson state
with the 4-velocity $v$ in the HQET,
$\overline{q}$ is the light-antiquark field, $h_v$ is the effective 
heavy-quark field,
and the heavy-light local operator in the LHS is renormalized at the scale $\mu$.
The decay constant of \eq{fmu} represents the quantitative content of 
the quark-antiquark valence component inside the $B$ meson in the heavy-quark limit, so that
$F(\mu)$ determines the normalization of the
valence Fock components in the $B$-meson wavefunction,
as well as of the amplitude for the exclusive $B$-meson decays.
We note that $F(\mu)$ is related to the physical decay constant $f_B$ as
\begin{eqnarray}
   f_B\sqrt{m_B}& =& F(\mu)\left[1+\frac{C_F\alpha_s}{4\pi}
   \left(3\ln\frac{m_b}{\mu}-2\right)+\ldots\right] 
\nonumber\\
&+&O(1/m_b),
\label{F-fB}
\end{eqnarray}
with the corresponding short-distance coefficient shown in the
parentheses to the one-loop accuracy, as well as with the $O(1/m_b)$
correction terms in the heavy-quark expansion.
The value of $f_B$ 
is now obtained rather precisely from lattice QCD calculations
as~\cite{Gamiz:2009ku} (see also \cite{Bernard,Blossier:2009hg})
\begin{equation}
f_B=0.195\pm 0.013~{\rm GeV} ,
\label{fBvalue}
\end{equation}
which is consistent~\cite{Schwartz:2009hv} with the results of
measurement of the branching fraction for $B \to \tau \nu$ decays
in the Belle~\cite{Hara:2010dk} and Babar~\cite{Aubert:2007bx} experiments.

We can also define 
the analogues 
of \eq{fmu}, which are associated with the higher Fock components inside the $B$ meson.
For the non-minimal parton configurations
with 
additional gluons,
the corresponding HQET parameters were introduced by Grozin and Neubert~\cite{GN} as
\begin{eqnarray}
&&\bra 0|\overline{q}\bfm{\alpha}\cdot g\bfm{E}\g_5h_v|\bar B(v)\ket=F(\mu)\lambda_E^2(\mu),
\label{lame}
\\
&&\bra 0|\overline{q}\bfm{\sigma}\cdot g\bfm{H}\g_5h_v|\bar B(v)\ket=iF(\mu)\lambda_H^2(\mu),
\label{lamh} 
\end{eqnarray}
in terms of the matrix elements in the $B$-meson rest frame with $v=(1, \bfm{0})$.
Here, the three-body quark-gluon operators are
associated with the chromoelectric and chromomagnetic fields,
$E^i=G^{0i}$ and $H^i=(-1/2)\varepsilon^{ijk}G^{jk}$,
with $G_{\mu\nu}= G_{\mu\nu}^a T^a$ and
$G_{\mu\nu}^a=\partial_\mu A^a_\nu-\partial_\nu A^a_\mu+gf^{abc}A^b_\mu A_\nu^c$
being the gluon field strength tensor. 
The values of $\lambda_{E,H}^2$ were also estimated in \cite{GN} using
QCD sum rules, as
\begin{eqnarray}
&&\lambda_E^2 (\mu) = 0.11 \pm 0.06~{\rm GeV}^2 ,
\nonumber\\
&&\lambda_H^2 (\mu) = 0.18 \pm 0.07~{\rm GeV}^2 ,
\label{lambdaEH}
\end{eqnarray}
at $\mu =1$~GeV.
Besides this rather rough estimate, there exists no other estimate at present.
In this paper, we present 
an extension of Grozin-Neubert's QCD sum rule calculation 
of $\lambda_{E,H}^2$, taking into account the higher-order perturbative and nonperturbative effects 
in QCD.
The main new ingredient in the present case
is that we calculate the relevant QCD radiative corrections, so that
we derive our sum rules to the order $\alpha_s$ accuracy.
We find that, only after including those new contributions, 
the perturbative as well as nonperturbative corrections
to the sum rules for $\lambda_{E,H}^2$ become under control.
We note that, also for the QCD sum-rule calculations of 
the decay constant (\ref{fmu}),
the order $\alpha_s$ radiative corrections 
produce the large and essential contributions
to yield the values consistent with \eq{fBvalue}~\cite{bagan,penin,neubert0}.

One might anticipate that the higher Fock components 
in the $B$ meson
would give rise to 
the ``higher-twist'' power
corrections to the hard exclusive amplitudes,
similarly as 
the roles played by 
the higher Fock components in
the light mesons $\pi$, $\rho$, etc.~\cite{zzc},
and thus the impact of a more accurate determination of $\lambda_{E,H}^2$ 
than \eq{lambdaEH} would be marginal.
Actually, however, it has been revealed
that the behaviors of the contributions
induced by the higher Fock components 
are quite different between the 
$B$-meson case and the light-meson case:
the presence of a heavy quark inside the $B$-meson causes the nonperturbative
quark-gluon interactions which induce 
the mixing of the effects corresponding to the different twist~\cite{GN,KKQT,GW}.
In particular, recently, it has been demonstrated that the $B$-meson ``light-cone distribution amplitudes'' 
to describe the valence Fock components participating 
in the hard exclusive processes~\cite{GN,KKQT,GW,braun,Lee:2005gza,DescotesGenon:2009hk}
are contaminated by 
the multiparticle Fock states, so that the contributions represented by 
the novel HQET parameters $\lambda_{E,H}^2$ of Eqs.~(\ref{lame}),~(\ref{lamh})
could strongly affect~\cite{Kawamura:2008vq} the amplitudes for the exclusive $B$-meson decays
at the {\em leading} power.
Indeed, the normalization of the so-called hard spectator interaction amplitude~\cite{Antonelli:2009ws}
could be modified by a factor two or more,
when varying the values of $\lambda_{E,H}^2$ in the uncertainty range 
of \eq{lambdaEH}~\cite{Kawamura:2008vq}.
Therefore, an improved estimate of $\lambda_{E,H}^2$ is desirable 
to have a better control of the hadronic uncertainty associated with the $B$ meson,
which is a major source of theoretical uncertainty 
in the calculations of the decay rates~\cite{Bell08,Antonelli:2009ws}.

The paper is organized as follows.
Section~\ref{sec2} is mainly introductory;
we set up a systematic formalism for the QCD sum rule calculation of
$\lambda_{E,H}^2$ in the HQET, as well as the decay constant $F$,
and apply it to reproduce the previous results for the sum rules of those HQET parameters.
We explain that the previous sum-rule estimate of $\lambda_{E,H}^2$ needs update including higher-order effects.
In Sec.~\ref{sec22}, we derive the new power corrections to the sum rules for $\lambda_{E,H}^2$,
due to the nonperturbative QCD condensates.
Section~\ref{sec3} is devoted to the calculation of the new order $\alpha_s$ corrections 
to the sum rules for $\lambda_{E,H}^2$.
Taking into account all these new contributions,
we present the final form of our sum rules for $\lambda_{E,H}^2$
in Sec.~\ref{sec5}. We explain the renormalization-group improvement of our sum rule formulas 
and perform their detailed numerical analysis to obtain a new estimate of $\lambda_{E,H}^2$.
We find that the new values of $\lambda_{E,H}^2$ are significantly modified from 
those of \eq{lambdaEH}.
Section~\ref{sec6} is reserved for conclusions.

\section{QCD sum rules in the HQET}
\label{sec2}

In this section we set up the framework convenient for treating
the perturbative as well as nonperturbative corrections to the suitable correlation functions in the HQET
for the QCD sum-rule calculations
of the $B$-meson matrix elements $\lambda_E$ and $\lambda_H$,
and also use it to demonstrate that the calculation of the leading effects 
reproduces the sum rules obtained previously by Grozin and Neubert~\cite{GN}.
The complete treatment including the new perturbative and nonperturbative corrections
is presented in the succeeding sections.

We consider the following correlation functions in the HQET
(the dependence on the renormalization scale $\mu$ is suppressed for simplicity):
\bey
&& \!\!\!\!\!\!\!\!\!\!\!\!\!\!\!\!
i\int d^4x e^{-i\omega v\cdot x}
\bra 0|T\left[\overline{q}(0)\Gamma_1  h_v(0)\ \h_v(x)\Gamma_2 q(x)
\right]|0\ket
\nonumber\\
&&
=-\frac{1}{2}\Tr\left[\Gamma_1\pp\Gamma_2\right]\Pi_F(\omega) ,
\label{corfun0}
\eey
\bey
&&\!\!\!\!\!\!\!\!\!\!\!\!
i\int d^4x e^{-i\omega v\cdot x}
\bra 0|T\left[\overline{q}(0)\Gamma_1 gG_{\mu\nu}(0) h_v(0)\ \h_v(x) \Gamma_2 q(x)
\right]
|0\ket
\nonumber\\
&&\!\!\!\!\!\!\!
= - \frac{1}{2}\Tr\left[\sig_{\mu\nu}\Gamma_1 \pp \Gamma_2 \right]\Pi_{3H}(\omega) 
\nonumber\\
&&
-  \frac{1}{2}
\Tr\left[(iv_\mu \gamma_\nu -iv_\nu \gamma_\mu)\Gamma_1 \pp \Gamma_2 \right]\Pi_{3S}(\omega),
\label{corfun}
\eey
where 
$P_+ =(1+\Slash{v})/2$
is the projector on the upper components of the heavy-quark spinor,
$\Gamma_1$ is an arbitrary gamma matrix, and we choose
$\Gamma_2=\g_5$ to construct the sum rules
for pseudoscalar $B$ meson.
(The case for
the vector meson $B^*$ can also be treated by choosing 
$\Gamma_2 = \gamma_\mu - v_\mu$ and yields exactly the same results
as in the pseudoscalar $B$-meson case
due to heavy-quark spin symmetry in the HQET.)
Eq.~(\ref{corfun0}) is the familiar correlator
between two heavy-light currents and the correlation function $\Pi_F(\omega)$
provides the sum rules
to evaluate the decay constant $F$.
On the other hand, the correlator (\ref{corfun}) between the two-body current
and 
the three-body current
involving the gluon field strength tensor
defines the correlation functions, $\Pi_{3H}(\omega)$ and $\Pi_{3S}(\omega)$,
corresponding to the two independent Lorentz structures; as we will show shortly,
$\Pi_{3H}(\omega)$ and $\Pi_{3S}(\omega)$ allow us to derive the sum rules
to evaluate $\lambda_{H}^2$ and the ``splitting'' $\lambda_{H}^2-\lambda_{E}^2$,
respectively (see Eqs.~(\ref{lame}), (\ref{lamh})).

In the general procedure of QCD sum rules, 
we evaluate the above correlation functions 
by  the operator product expansion (OPE) in the unphysical region $-\omega \gg \Lambda_{\rm QCD}$ 
on one hand and 
express those correlation functions in terms of the properties
(masses and matrix elements) associated with the physical states participating in the
spectra at $\omega >0$
on the other hand;
we relate these two descriptions  
exploiting the analyticity properties of the correlation functions,
which are embodied by 
the corresponding dispersion relations.
The dispersion relation satisfied by $\Pi_{F}(\omega)$ of Eq.~(\ref{corfun0})
is well-known, and the 
dispersion relations of similar type are obeyed also by $\Pi_{3H}(\omega)$ and $\Pi_{3S}(\omega)$
of (\ref{corfun}); namely, for $X=F, 3H, 3S$, 
\ben 
\Pi_X(\omega)=\frac{1}{\pi}\int^\infty_{0}d\omega'\frac{{\rm Im} \Pi_X(\omega')}{\omega'-\omega-i0},
\label{disp}
\een
up to the appropriate subtraction terms that are polynomial in $\omega$.
Indeed, these relations can be demonstrated 
by inserting a complete set of states between the two currents in the corresponding correlators:
the LHS of Eq.~(\ref{corfun0}) yields (in the $B$-meson rest frame)
\bey
&&\!\!\!\!\!\!\!\!\!\!\!\!\!\!\!\!\!\!\!\!\!\!
\frac{1}{2\left(\bar\Lambda -\omega- i0\right)}
\bra 0|\overline{q}(0) \Gamma_1 h_v(0)|\bar B(v)\ket
\cr&&\;\;\;\;\times
\bra \bar B(v)|\overline{h}_v(0)\Gamma_2 q(0)|0\ket
+\cdots,
\label{cf-B}
\eey
and the LHS of Eq.~(\ref{corfun}) gives the similar
result with $\Gamma_1$ replaced by $\Gamma_1 gG_{\mu\nu}(0)$.
Here, a pole arises at $\bar\Lambda=m_B -m_b$, 
the usual effective mass of the $B$ meson,
accompanying the 
matrix elements that are parameterized by
the corresponding HQET parameters (renormalized at the scale $\mu$) as
\bey
&&\bra 0|\overline{q}(0)\Gamma_1 h_v(0)|\bar B(v)\ket=-\frac{i}{2}F(\mu)\Tr\left[\Gamma_1\pp\g_5\right],
\label{mat2}\\
&&\bra 0|\overline{q}(0)\Gamma_1 gG_{\mu\nu}(0)h_v(0)|\bar B(v)\ket
\cr&&=
-\frac{i}{6}F(\mu)\left\{\lambda_H^2(\mu)\Tr\left[\Gamma_1\pp\g_5\sigma_{\mu\nu}\right]
\right.\cr&&\left.
+[\lambda_H^2(\mu)-\lambda_E^2(\mu)]
\Tr\left[\Gamma_1\pp\g_5(iv_\mu \gamma_\nu -iv_\nu \gamma_\mu)
\right]\right\},
\label{mat3}
\eey
and the ellipses stand for the similar pole contributions of higher resonances and the continuum contributions. 
Combining the results with Eqs.~(\ref{corfun0}), (\ref{corfun}), one finds,
\bey
&& \!\!\!\!\!\!\!\!\!\!\!
\Pi_F(\omega) 
= \frac{F^2(\mu)}{2}\frac{1}{\bar\Lambda -\omega- i0}+\cdots,
\cr
&&\!\!\!\!\!\!\!\!\!\!\!\!
\Pi_{3H}(\omega)
=\frac{F^2(\mu)}{6}\lambda_H^2(\mu)\frac{1}{\bar\Lambda -\omega- i0}+\cdots,
\cr
&&\!\!\!\!\!\!\!\!\!\!\!\!
\Pi_{3S}(\omega) 
=\frac{F^2(\mu)}{6}\left[\lambda_H^2(\mu)-\lambda_E^2(\mu)\right]
\frac{1}{\bar\Lambda -\omega- i0}+\cdots,
\label{disp2}
\eey
which hold in any frame as well as in the rest frame,
and show that the contribution of the $B$ meson to the spectral
functions in Eq.~(\ref{disp}) is completely expressed by the relevant HQET parameters.
Calculating the LHS of Eqs.~(\ref{corfun0}), (\ref{corfun}) based on the OPE 
and matching the results with the formulas in \eq{disp2},
we obtain the sum rules associated with those HQET parameters.

We follow the standard procedure to construct the corresponding QCD sum rules: we apply
the Borel-transformation operator, defined by
\bey
\hat{B}_M\equiv\lim_{\stackrel{n\rightarrow\infty,-\omega\rightarrow\infty}{M=-\omega/n\ {\rm fixed}}}
\frac{\omega^n}{\Gamma(n)}\left(-\frac{d}{d\omega}\right)^n,
\label{borel}
\eey
to the relevant correlation functions obeying the dispersion relation (\ref{disp}), (\ref{disp2}).
This transformation 
introduces the Borel parameter $M$ instead of the external energy $\omega$
as
\bey
\hat{B}_M\Pi_X(\omega)=\frac{1}{M}\int^\infty_{0}d\omega'
e^{-\omega'/M}
\frac{1}{\pi}{\rm Im}\Pi_X(\omega'),
\label{BorelSR}
\eey
and eliminates the subtraction terms. 
\eq{borel} implies
that the power-correction terms associated with the higher dimensional operators in the OPE 
are factorially ($\sim 1/n!$) suppressed, improving the convergence of the series
and, simultaneously, 
\eq{BorelSR} indicates that the contributions of higher resonances and continuum 
are exponentially suppressed compared with that of the lowest-lying state,
minimizing the dependence on the contributions of the 
excited states.
Employing quark-hadron duality,
we approximate, as usual, 
those excited-state contributions 
to the spectral function in \eq{BorelSR} by the 
continuum contribution which is
based on the OPE result and starts from
the \lq\lq continuum threshold" 
$\omega_{\rm th}$; namely, we use
\bey
\frac{1}{\pi}{\rm Im}\Pi_F(\omega)
&=&\frac{1}{2}F^2(\mu) \delta(\omega-\bar\Lambda)
\cr&&
+\frac{1}{\pi}{\rm Im}\Pi_F^{\rm OPE}(\omega)\theta(\omega-\omega_{\rm th}),
\label{spectralansatzF}
\eey
with the correlation function $\Pi_F^{\rm OPE}(\omega)$ 
calculated in the OPE, and the similar form for 
$(1/\pi){\rm Im}\Pi_{X}(\omega)$ ($X=3H, 3S$)
with the corresponding OPE result, $\Pi_{X}^{\rm OPE}(\omega)$. 
Then, we obtain the sum rules,
\bey
&& \!\!\!\!\!\!\!\!\!\!\!\!\!\!\!
F^2(\mu) e^{-\bar\Lambda/M} = 
2\int_0^{\omega_{\rm th}} \!\!d\omega e^{-\omega/M}\frac{1}{\pi}{\rm Im}\Pi_F^{\rm OPE}(\omega),
\label{BorelSR0}\\
&& \!\!\!\!\!\!\!\!\!\!\!\!\!\!\!
F^2(\mu)\lambda_H^2(\mu) e^{-\bar\Lambda/M}=
6 \int_0^{\omega_{\rm th}} \!\!d\omega e^{-\omega/M}\frac{1}{\pi}{\rm Im}\Pi_{3H}^{\rm OPE}(\omega),
\label{BorelSR1}\\
&& \!\!\!\!\!\!\!\!\!\!\!\!\!\!\!
F^2(\mu)\left[ \lambda_H^2(\mu)-\lambda_E^2(\mu)\right] e^{-\bar\Lambda/M}
\cr &&
=6 \int_0^{\omega_{\rm th}} d\omega e^{-\omega/M}\frac{1}{\pi}{\rm Im}\Pi_{3S}^{\rm OPE}(\omega),
\label{BorelSR2}
\eey
and, taking the ratios of Eqs.~(\ref{BorelSR1}) and (\ref{BorelSR2}) with \eq{BorelSR0}
to cancel the factor $F^2(\mu)$,
one can evaluate $\lambda_{H}^2(\mu)$ and 
$\lambda_H^2(\mu)-\lambda_E^2(\mu)$, respectively, based on the sum rules.

Now the remaining task is to calculate $\Pi_X^{\rm OPE}(\omega)$,
to be substituted into the RHS of Eqs.~(\ref{BorelSR0})-(\ref{BorelSR2}).
Carrying out the OPE of the corresponding 
correlation functions for the region $-\omega \gg \Lambda_{\rm QCD}$, 
the results generically take the form,
\bey
\Pi_{X}^{\rm OPE}(\omega)&=&C^{X}_I(\omega)+C^{X}_q(\omega)\qcon
+C^{X}_G(\omega)\gcn
\cr&&+C^{X}_\sigma(\omega)\mixqn
+\cdots,
\label{OPEgeneral}
\eey
with $X=F, 3H, 3S$, where 
$\qcon\equiv \bra 0| \overline{q}q |0 \ket$, $\gcn\equiv \bra 0| (G_{\mu\nu}^a)^2|0 \ket$, and
$\mixqn\equiv \bra 0|  \overline{q} gG^{\mu\nu} \sigma_{\mu \nu}q |0 \ket$
are the quark condensate,
the gluon condensate,
and the quark-gluon-mixed condensate, respectively,
as the vacuum expectation values of the dimension-3, -4, and -5
local operators, and are associated with the corresponding  
Wilson coefficients $C^X_k(\omega)$ with $k=q, G$, and $\sigma$;
an increase in dimension of the operators implies extra powers
of $1/\omega$ for the corresponding Wilson coefficients,
and the ellipses in \eq{OPEgeneral}
denote the terms 
with the operators of dimension $d \ge 6$.
$C^X_I(\omega)$, associated with the unit operator, 
coincides with the purely perturbative contribution 
to $\Pi_X(\omega)$.
The condensates as well as the coefficient functions in general depend on 
the renormalization scale $\mu$.

\begin{figure}
\includegraphics[width=7cm,keepaspectratio]{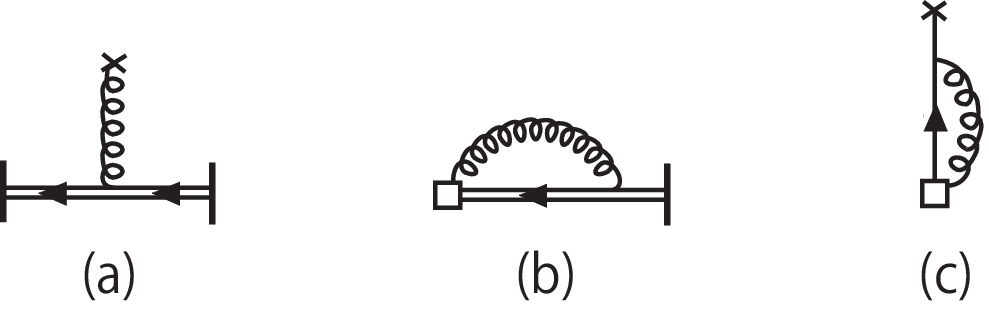}
\caption{The vanishing subdiagrams. (a) is associated with classical
background gluon field of \eq{fs}. (b) and (c) have a loop involving 
quantum gluon field emanating from 
the field strength tensor in the three-body current.
The external lines with a bar at their end are 
amputated.}
\label{subdiagram}
\end{figure}

For the correlation function (\ref{corfun0})
with the two-body currents,
the OPE can be 
derived in a standard way
and the Wilson coefficients appearing in \eq{OPEgeneral}
with $X=F$ are obtained~\cite{shuryak,neubert,bagan,broad,penin} 
by evaluating the familiar Feynman diagrams,
which involve the heavy-quark propagator 
in a background 
gluon field $A_\mu$,
\begin{eqnarray}
\wick[u]{2}{<1{h_v(0)} >1{\h_v(x)}}
&=&
\theta(-v\cdot x)\delta^{(D-1)}\left(x_\perp\right) P_+ 
\nonumber\\
&\times&\!\!\!
{\cal P}\exp\left(ig \int_{v\cdot x}^0 ds v\cdot A(sv)\right),
\label{hprop}
\end{eqnarray}
for the case of the $D$ dimensions, where $x_\perp^{\mu}=x^\mu -(v\cdot x)v^\mu$, and
the factor in the second line, i.e.,
the straight Wilson line along the velocity $v$ 
with the path-ordering operator ${\cal P}$, allows us to
organize 
the interactions with the gluon field $A_\mu$ exactly
in the HQET.
Thus, 
it is convenient to use the Fock-Schwinger gauge, $x^\mu A_\mu(x)=0$,
for the background gluon field, so that the heavy quark does not
interact with the nonperturbative gluons
in the calculation for power corrections to
the correlation function (\ref{corfun0}) (see Fig.~\ref{subdiagram}(a) below).
In this case, 
it is also well-known
that
we have
very useful relations~\cite{novikov}:
for the classical background gluon field,
\begin{equation}
A_\mu(x) = \int_0^1 du u x^\nu G_{\nu \mu}(ux),
\label{fs}
\end{equation}
and, for the light-quark propagator,
\bey
&&\!\!\! \!\!\! \!\!    \!\!\! \!\!    \!\!\! \!\!   
\wick[u]{2}{<1{q(x)} >1{\overline{q}(0)}}
=
\frac{i \Gamma\left(\frac{D}{2}\right)\Slash{x}}
{2\pi^{\frac{D}{2}} \left(-x^2+i0 \right)^{\frac{D}{2}}} 
\nonumber\\  
&&\!\!\! \!\!\! \!\!   
+ 
\frac{i\Gamma \left( {{\frac{D}{2}}- 1} \right)}
{32\pi^{\frac{D}{2}}\left(-x^2+i0 \right)^{{\frac{D}{2}}- 1}}
\left\{\Slash{x} , \sigma^{\mu \nu} \right\}
gG_{\mu \nu} (0)
+\cdots ,
\label{qprop}
\eey
with the ellipses denoting the terms associated with operators
of dimension $d \ge 3$;
it is worth noting that the term associated with the dimension-4 
operator $G^2$ is absent from the ellipses.
We do not give the details of the calculation of 
the Wilson coefficients here
but, for later convenience, sketch the 
relevant steps:
we decompose the quark as well as gluon fields into the ``quantum'' and ``classical'' parts;
the quantum parts are contracted to yield the propagators in the classical
background fields, like Eqs.~(\ref{hprop}), (\ref{qprop}),
while the classical parts satisfy the classical equations of motion, 
$\Slash{D}q=0$, $v\cdot D h_v=0$, 
and $D^\nu G_{\mu \nu}^a = g \sum_{q'}\overline{q'}\gamma_\mu T^a q'$
with the summation over all quark flavors.
For the matching at the leading accuracy in $\alpha_s$,
the correlator in the LHS of \eq{corfun0} is evaluated as
\bey
T\left[\overline{q}(0)\Gamma_1  h_v(0)\ \overline{h}_v(x)\Gamma_2 q(x)
\right]
&& \!\!\!\!\!
= \wick[u]{32}{<2{\overline{q}(0)} \Gamma_1  <1{h_v(0)} >1{\overline{h}_v(x)}\Gamma_2 
>2{q(x)}}
\nonumber\\
&& \!\!\!\!\!\!\!\!\!\!\!\!\!\!\!\!\!\!\!\!
+ 
\overline{q}(0) \Gamma_1  \wick[u]{2}{<1{h_v(0)} >1{\overline{h}_v(x)}}\Gamma_2 q(x) ,
\label{corfun0wick}
\eey 
and, substituting Eqs.~(\ref{hprop})-(\ref{qprop}) into the first term
in the RHS, 
we immediately obtain $C^{F}_{I}$ in \eq{OPEgeneral} 
as the purely perturbative contribution
and also find that $C^{F}_{G}$ vanishes up to the corrections of $O(\alpha_s^2)$,
as a direct consequence of the absence of the operator $G^2$ in \eq{qprop} 
as noted above.
On the other hand, the vacuum expectation value of the second term 
in the RHS of \eq{corfun0wick} yields, 
\bey
&&
\theta(-v\cdot x)\delta^{(3)}\left(x_\perp\right) 
\bra 0|\overline{q}(0) \Gamma_1 P_+ \Gamma_2 q(x) |0\ket ,
\nonumber\\
&&=\theta(-v\cdot x)\delta^{(3)}\left(x_\perp\right) \left[
\bra 0| \overline{q}(0) \Gamma_1 P_+ \Gamma_2 q(0) |0\ket \right.
\nonumber\\
&&+ x^\mu \bra 0| \overline{q}(0) \Gamma_1 P_+ \Gamma_2 D_\mu q(0) |0\ket
\nonumber\\
&& 
+ \frac{1}{2}\left. 
x^\mu x^\nu \bra 0| \overline{q}(0) \Gamma_1 P_+ \Gamma_2 D_\mu D_\nu q(0) |0\ket
+\cdots
\right]
\nonumber\\
&&=\theta(-v\cdot x)\delta^{(3)}\left(x_\perp\right)
\Tr\left[\Gamma_1\pp\Gamma_2\right]
\nonumber\\
&&\times
\frac{1}{4} \left[\qcon +\frac{1}{16}x^2 \mixqn +\cdots 
\right],
\label{corfun0wick2}
\eey
where we used the equations of motion $\Slash{D}q=0$ in the last equality,
and this allows us to obtain $C^F_q$ as well as $C^F_\sigma$.
As the result, the relevant Wilson coefficients
read~\cite{shuryak,neubert,bagan} 
\bey
C^{F}_{I}(\omega)&=&- \frac{N_c}{2 \pi^2} \omega^2
\ln\frac{-\omega}{\mu},\cr
C^{F}_{q}(\omega)&=&\frac{1}{2\omega},\cr
C^{F}_{G}(\omega)&=&0 ,\cr
C^{F}_{\sigma}(\omega)&=&-\frac{1}{16\omega^3} ,
\label{WCFR<5}
\eey
up to the terms polynomial in $\omega$.
We calculate the discontinuities of \eq{OPEgeneral} with these coefficients,
across the cut along the line $\omega >0$ in the complex $\omega$ plane,
and substitute the results into the RHS of \eq{BorelSR0}.
This yields the sum rule,
\bey
F^2(\mu) e^{-\bar\Lambda/M}\!\!\!\!\!\!
&&=\frac{2}{\pi^2}N_cM^3 W^{(2)}(\frac{\omega_{\rm th}}{M})
-\qcon \cr&&
+\frac{1}{16M^2}\mixqn ,
\label{fsr0}
\eey
where 
the function,
\begin{equation}
W^{(m)}(x)\equiv 1-\sum_{k=0}^m\frac{x^k}{k!}e^{-x} ,
\label{Wm}
\end{equation}
arises from the integral over the duality region, $0<\omega <\omega_{\rm th}$,
in \eq{BorelSR0}.
This sum rule can be used for a leading estimate of 
the decay constant $F(\mu)$.
\begin{figure}
\includegraphics[width=5cm,keepaspectratio]{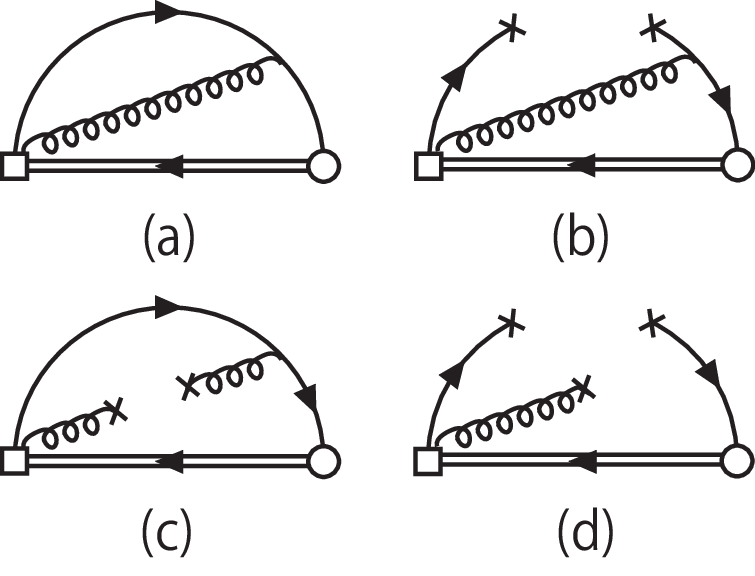}
\caption{Nonvanishing diagrams for the OPE 
of the correlation function (\ref{corfun})
in the Fock-Schwinger gauge.
The double line denotes the propagator of the effective heavy quark, and 
the circle and box represent the two- and three-body currents, respectively.
The four diagrams (a)-(d)
generate, respectively, the first four terms of (\ref{OPEgeneral})
with $X=3H, 3S$.
}
\label{treediagram}
\end{figure}

The sum rules for a leading estimate of $\lambda_{E,H}^2$ 
can be derived similarly.
The corresponding calculation 
was performed
by Grozin and Neubert~\cite{GN}
with two particular choices for the gamma matrix $\Gamma_1$ of (\ref{corfun}),
which make
the corresponding three-body currents 
coincide with the chromoelectric and chromomagnetic operators
in the LHS of Eqs.~(\ref{lame}) and (\ref{lamh}).
For later convenience and for a cross-check of Grozin-Neubert's result,
we here perform the corresponding calculation 
for arbitrary gamma matrix $\Gamma_1$, and summarize the procedures and the 
results.
The correlator between the two-body and three-body currents
in the LHS of \eq{corfun} is evaluated as
\bey
&&T\left[\overline{q}(0)\Gamma_1 gG_{\mu\nu}(0)  h_v(0)\ \overline{h}_v(x)\Gamma_2 q(x)
\right]
\nonumber\\
&&= 
\wick[u]{32}{<2{\overline{q}(0)} \Gamma_1 gG_{\mu\nu}(0)  
 <1{h_v(0)} >1{\overline{h}_v(x)}\Gamma_2 
>2{q(x)}}
\nonumber\\
&& 
+ 
\overline{q}(0) \Gamma_1 gG_{\mu\nu}(0)   \wick[u]{2}{<1{h_v(0)} >1{\overline{h}_v(x)}}\Gamma_2 q(x) 
+ \cdots .
\label{corfunwick}
\eey
By contrast to the above case leading to the results (\ref{WCFR<5}),
the extra gluons emanating from 
the three-body current participate in the present case.
Those extra gluons can interact with the light quark and such contributions
require the participation of the additional quark-gluon coupling in perturbation theory,
so as to form the propagator,
\bey
&&\!\!\! \!\!\! 
\wick[u]{2}{<1{G_{\mu \nu}^a(0)} >1{A_{\lambda}^b(z)}}
=
\frac{\Gamma\left(\frac{D}{2}\right) \delta^{ab}}
{2\pi^{\frac{D}{2}} \left(-z^2+i0 \right)^{\frac{D}{2}}} 
\left( g_{\nu \lambda}z_\mu-g_{\mu \lambda}z_\nu \right)
\nonumber\\  
&&\;\;
+ 
\frac{\Gamma \left( {{\frac{D}{2}}- 1} \right)f^{abc}}
{8\pi^{\frac{D}{2}}\left(-z^2+i0 \right)^{{\frac{D}{2}}- 1}}
\left(gG_{\mu \rho}^c (0)z^\rho g_{\nu\lambda}-2gG_{\mu \lambda}^c (0)z_\nu 
\right.
\nonumber\\
&&\;\;
\left.
- gG_{\nu \rho}^c (0)z^\rho g_{\mu\lambda} +2gG_{\nu \lambda}^c (0)z_\mu \right)
+\cdots ,
\label{gprop}
\eey
for the case of the $D$ dimensions
and using the Feynman gauge for the quantum part of the gluon field, with 
the ellipses denoting the terms associated with operators
of dimension $d \ge 3$.
The ellipses in \eq{corfunwick}
stand for the terms of this type at the leading accuracy in $\alpha_s$,
which are induced by the first term in the RHS of \eq{gprop},
and the corresponding 
nonvanishing contributions are
%
represented by the Feynman
diagrams (a) and (b) in Fig.~\ref{treediagram};
note that, by explicit calculation, 
the subdiagrams (b), (c) in Fig.~\ref{subdiagram}
vanish, reflecting that
a physical gluon represented by the field strength tensor
does not interact with the heavy quark, nor is absorbed into 
a single quark on the mass shell. (The contributions
induced by the second term of \eq{gprop} will be discussed in Sec.~\ref{sec3}.)
Decomposing those contributions from Figs.~\ref{treediagram}~(a) and (b)
into independent Lorentz structures, as in the RHS of \eq{corfun},
we obtain the Wilson coefficients 
$C_I^X$ and $C_q^X$, respectively, in the OPE (\ref{OPEgeneral}) with $X=3H, 3S$.

Similarly, it is straightforward to see that 
the vacuum expectation value of the first term in the RHS of
\eq{corfunwick} yields $C_G^{3H,3S}$,
by combining
the field strength tensor of the first term of \eq{corfunwick}
with that from 
the second term of the quark propagator~(\ref{qprop}),
corresponding to the diagram (c)
in Fig.~\ref{treediagram}.
On the other hand, 
the contribution of
the second term in the RHS of
\eq{corfunwick} can be evaluated similarly as \eq{corfun0wick2},
and we obtain, 
\bey
&& \!\!\!\!\!\!\!
\theta(-v\cdot x)\delta^{(3)}\left(x_\perp\right)
\bra 0|\overline{q}(0) gG_{\mu\nu}(0) \Gamma_1 P_+ \Gamma_2 q(x)
|0\ket
\nonumber\\
&&\!\!\!\!\!\!\!
=
\theta(-v\cdot x)\delta^{(3)}\left(x_\perp\right)
\nonumber\\
&&\times\left(\frac{1}{48}\Tr\left[\Gamma_1\pp\Gamma_2 
\sigma_{\mu\nu} \right]\mixqn
+ \cdots\right),
\label{corfunwick2}
\eey
with the term represented by the diagram~(d) 
in Fig.~\ref{treediagram} and the ellipses denoting the contributions associated with the operators
of dimension $d \ge 6$, so that we can calculate $C_\sigma^{3H,3S}$
using the former contribution.
Collecting the results from the diagrams (a)-(d) in \fig{treediagram},
we obtain the corresponding Wilson coefficients as
\bey
C^{3H}_I(\omega)&=&\frac{1}{18 \pi^2}N_cC_F\frac{\alpha_s}{4\pi}\omega^4
\ln\frac{-\omega}{\mu},\cr
C^{3H}_q(\omega)&=&C_F\frac{\alpha_s}{2\pi}\omega\ln\frac{-\omega}{\mu},\cr
C^{3H}_G(\omega)&=&-\frac{\alpha_s}{24\pi} \ln\frac{-\omega}{\mu},\cr
C^{3H}_\sigma (\omega)&=&\frac{1}{24\omega},
\label{WC3H<5}
\eey
and 
\bey
C^{3S}_I(\omega)&=&-\frac{1}{18\pi^2}N_cC_F\frac{\alpha_s}{4\pi}\omega^4
\ln\frac{-\omega}{\mu},\cr
C^{3S}_q(\omega)&=&C_F\frac{\alpha_s}{2\pi}\omega\ln\frac{-\omega}{\mu},\cr
C^{3S}_G(\omega)&=&-\frac{\alpha_s}{24\pi}\ln\frac{-\omega}{\mu},\cr
C^{3S}_\sigma(\omega)&=&0,
\label{WC3S<5}
\eey
up to the terms polynomial in $\omega$, which are irrelevant for the present purpose.
Here, $C_F = (N_c^2-1)/(2N_c)$, and we 
note 
that the dimension-5 mixed condensate
does not contribute to $\Pi_{3S}^{\rm OPE}$.
Substituting (\ref{OPEgeneral}) with the coefficient functions
(\ref{WC3H<5}) and (\ref{WC3S<5}) into the RHS of
Eqs.~(\ref{BorelSR1}) and (\ref{BorelSR2}), respectively,
we obtain 
\bey
F^2(\mu)\lambda_H^2(\mu) e^{-\bar\Lambda/M}&&
\!\!\!\!\!\!
=
-\frac{2\alpha_s}{\pi^3}
N_cC_FM^5W^{(4)}(\frac{\omega_{\rm th}}{M})
\nonumber\\
&&\!\!\!\!\!\!\!\!\!
- \frac{3\alpha_s}{\pi}C_FM^2\qcon W^{(1)}(\frac{\omega_{\rm th}}{M}) 
\nonumber\\
&& \!\!\!\!\!\!\!\!\!\!\!\!\!\!\!\!\!\!\!\!\!\!\!\!\!\!\!\!\!\!\!\!
+\frac{M}{4} 
\gc W^{(0)}(\frac{\omega_{\rm th}}{M}) 
-\frac{1}{4} \mixqn ,
\label{treeSRformagnetic}
\eey
and
\bey
\lefteqn{F^2(\mu)\left[\lambda_H^2(\mu)-\lambda_E^2(\mu)\right]e^{-\bar\Lambda/M}}
\nonumber\\
&& \;\;\;\;\;\;\;\;\;\;\;\;
=\frac{2\alpha_s}{\pi^3}
N_cC_FM^5W^{(4)}(\frac{\omega_{\rm th}}{M})
\nonumber\\
&& \;\;\;\;\;\;\;\;\;\;\;\;
-\frac{3\alpha_s}{\pi}C_FM^2 \qcon W^{(1)}(\frac{\omega_{\rm th}}{M})
\nonumber\\
&& \;\;\;\;\;\;\;\;\;\;\;\;
+ \frac{M}{4} \gc W^{(0)}(\frac{\omega_{\rm th}}{M}) ,
\label{treeSRforsplitting}
\eey
where, as usual, the factor $\alpha_s/\pi$ from the coefficient functions
$C_G^{3H,3S}$ of Eqs.~(\ref{WC3H<5}), (\ref{WC3S<5})
is combined with the gluon condensate.
This set of the sum rules reproduces
Grozin-Neubert's sum rule formulas~\cite{GN} 
for $\lambda_H^2(\mu)$ and $\lambda_E^2(\mu)$.

In numerical evaluations throughout this paper, 
we use the standard values for the input parameters 
collected in Table~\ref{parameter}.
These values have been used in, e.g., a recent QCD sum rule calculation 
for the $B$-meson light-cone distribution amplitude~\cite{braun},
and are consistent with the values 
used in 
\cite{GN}; note that
the values of the condensates have been extracted at the $\sim 30$\% level accuracy~\cite{penin}.
$\alpha_s(1~{\rm GeV})=0.4$ was used in \cite{GN},
but we use the value $\alpha_s(1~{\rm GeV})=0.47$ which is consistent with the world average.

\begin{figure}
\includegraphics[width=8.5cm,keepaspectratio]{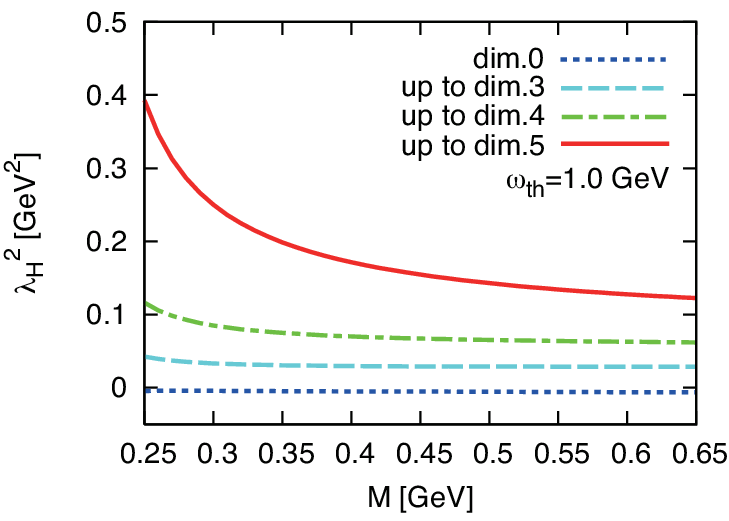}
\caption{The separate contributions to the Borel sum rule for $\lambda_H^2$
with $\mu=1$~GeV
and $\omega_{\rm th}=1.0\gev$, using \eq{treeSRformagnetic} divided by \eq{fsr0}. 
The contributions from each term in the RHS of \eq{treeSRformagnetic},
organized according to the dimension of the associated operators, are shown.
}
\label{ope-conv-H}
\end{figure}

\begin{figure}
\includegraphics[width=8.5cm,keepaspectratio]{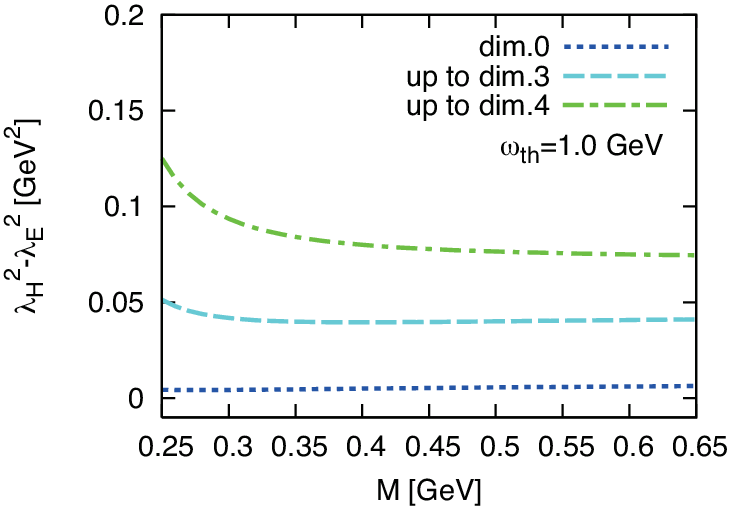}
\caption{Same as Fig.~\ref{ope-conv-H}, but for $\lambda_H^2-\lambda_E^2$, using \eq{treeSRforsplitting} divided by \eq{fsr0}. 
}
\label{ope-conv-S}
\end{figure}

\begin{figure}
\includegraphics[width=8.5cm,keepaspectratio]{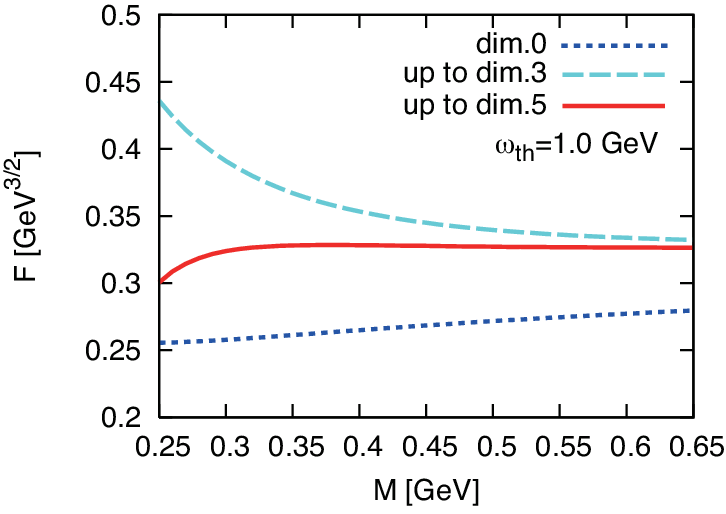}
\caption{The separate contributions to the Borel sum rule 
for $F$, using \eq{fsr0}
with $\mu=1$~GeV,
$\omega_{\rm th}=1.0\gev$, and $\bar\Lambda=0.55\gev$.
The contributions from each term in the RHS of \eq{fsr0},
organized according to the dimension of the associated operators, are shown.
}
\label{ope-conv-F}
\end{figure}

\begin{table}
\begin{center}
\begin{tabular}{cc}
\hline
\hline
Parameter & Value\\
\hline
$\qcon$ & $(-0.24\pm 0.02)^3~\gev^3$\\
$\gc$ & $(0.012\pm 0.006)~\gev^4$\\
$\mixqn/\qcon $ & $(0.8\pm 0.2)~\gev^2$\\
\hline
\hline
\end{tabular}
\end{center}
\caption{Input values of the vacuum condensates
at the normalization point $\mu= 1~\gev$.}
\label{parameter}
\end{table}%
\begin{figure}
\includegraphics[width=8.5cm,keepaspectratio]{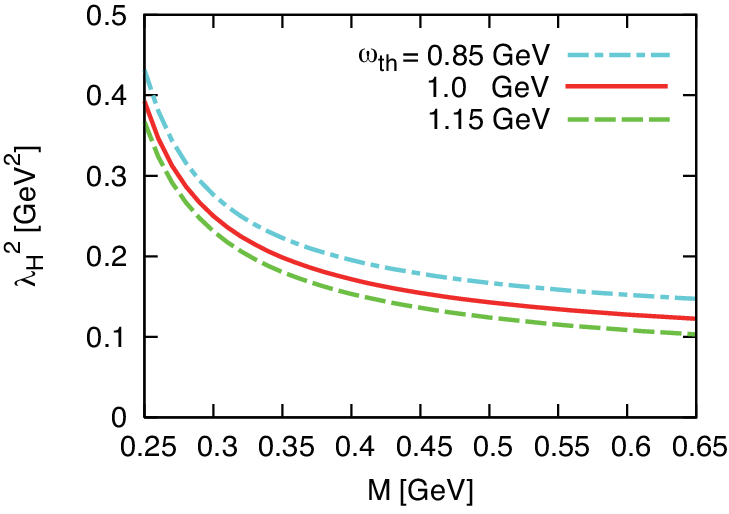}
\caption{Borel sum rule for $\lambda_H^2$ at $\mu=1$~GeV,
using \eq{treeSRformagnetic} divided by \eq{fsr0}.
}
\label{mag-tree}
\end{figure}
\begin{figure}
\includegraphics[width=8.5cm,keepaspectratio]{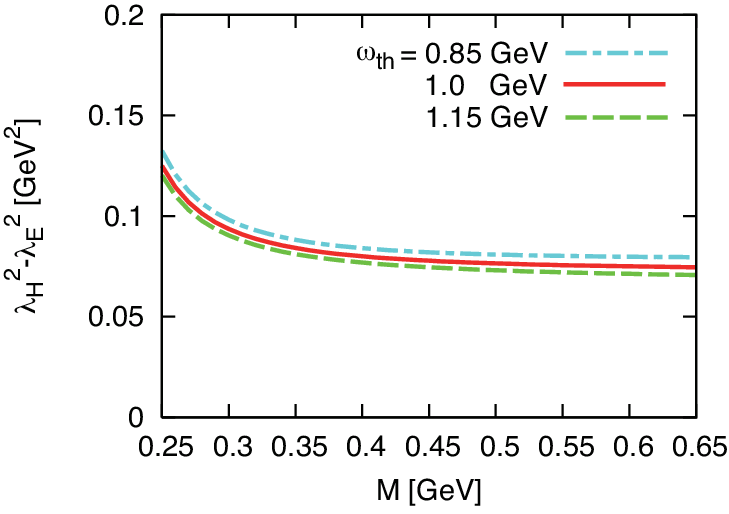}
\caption{Borel sum rule for $\lambda_H^2-\lambda_E^2$ at $\mu=1$~GeV,
using \eq{treeSRforsplitting} divided by \eq{fsr0}.
}
\label{split-tree}
\end{figure}
In Figs.~\ref{mag-tree} and \ref{split-tree}, we plot $\lambda_H^2(\mu)$ 
and $\lambda_H^2(\mu)-\lambda_E^2(\mu)$ at $\mu=1$~GeV as functions of 
$M$, 
obtained by taking 
the ratios of \eqs{treeSRformagnetic}{treeSRforsplitting}, respectively, with \eq{fsr0}.
One finds that the values of $\lambda_H^2$ are larger than those of
the splitting $\lambda_H^2-\lambda_E^2$,
and, in particular, that 
the curves for the former
show sizeable dependence on the parameter $M$.
Indeed, this considerable variation of $\lambda_H^2$ 
for $0.3~{\rm GeV} \lesssim M \lesssim 0.5~{\rm GeV}$,
which was taken in \cite{GN} as the ``stability window''
for the sum rule (\ref{fsr0}),
is responsible for the large errors in \eq{lambdaEH}.
Such poor stability is known to
be a common feature to the sum rules
for matrix elements of operators with high dimension~\cite{zzc},
because the corresponding sum rules are dominated at small $M$
by the condensates of high dimension.
For the present case with the dimension-5 operators in 
Eqs.~(\ref{lame}), (\ref{lamh}),
the behavior of 
the sum rule (\ref{treeSRformagnetic}) 
for $0.3~{\rm GeV} \lesssim M \lesssim 0.5~{\rm GeV}$
is mainly determined by the term with
the quark-gluon-mixed condensate $\mixqn$: 
this is demonstrated 
in \fig{ope-conv-H}, which shows the separate contributions from each term in the RHS
of (\ref{treeSRformagnetic}), organized according to the dimension 
of the associated local operators.
On the other hand,
the term with $\mixqn$ is absent from the sum rule (\ref{treeSRforsplitting}):
this sum rule yields a rather
stable behavior for $\lambda_H^2-\lambda_E^2$ as shown in \fig{split-tree},
but 
the separate contributions of each term in the RHS of \eq{treeSRforsplitting},
shown in \fig{ope-conv-S}, indicate that the nonperturbative corrections
do not decrease
for increasing dimension $d=0, 3$, and $4$ of the associated operators,
similarly as in \fig{ope-conv-H}.
These characteristic behaviors in Figs.~\ref{mag-tree} and~\ref{split-tree} are
in contrast to the case of the decay constant,
for which the 
separate contributions to the sum rule (\ref{fsr0}) are
plotted in \fig{ope-conv-F} with the value $\bar{\Lambda}=0.55$~GeV~\cite{GN}.
These results suggest good convergence of the OPE (\ref{OPEgeneral}) 
for $X=F$,
with the operators of dimension $d\le 5$,
while the convergence of \eq{OPEgeneral} for $X=3H$, $3S$, at the same level 
of accuracy, is questionable.
Therefore, 
we will calculate
the nonperturbative corrections to \eq{OPEgeneral}, associated with 
the dimension-6 operators, 
and evaluate the corresponding modifications to the sum rules
(\ref{treeSRformagnetic}) and (\ref{treeSRforsplitting}),
as well as to Eq.~(\ref{fsr0}),
in the next section.

The results for the relevant Wilson coefficients, Eqs.~(\ref{WC3H<5})
and (\ref{WC3S<5}),
show that only $C_\sigma^{3H}$, associated with the dimension-5 quark-gluon-mixed
condensate, is of $O(\alpha_s^0)$,
while all the other coefficients are of $O(\alpha_s)$.
This is again in contrast to the case of the decay constant,
for which all the nonzero
coefficients in \eq{WCFR<5}
are of $O(\alpha_s^0)$.
Combined with the behaviors in Figs.~\ref{ope-conv-H}-\ref{ope-conv-F} discussed above,
in particular,
with those indicating 
the dominance of the term associated with the dimension-5 operator in \eq{treeSRformagnetic},
it is desirable to calculate 
the $O(\alpha_s)$ correction to this term induced by the next-to-leading order (NLO)
correction to the 
Wilson coefficient $C_\sigma^{3H}$ of \eq{WC3H<5}.
It should be also clarified whether 
$C_\sigma^{3S}$ of \eq{WC3S<5}
receive the $O(\alpha_s)$ effects. 
These new $O(\alpha_s)$ contributions
may give the effects 
comparable with 
the other $O(\alpha_s)$ contributions
displayed in Figs.~\ref{ope-conv-H}, \ref{ope-conv-S}.
We will work out the one-loop matching 
to calculate the corresponding Wilson coefficients at $O(\alpha_s)$  
in Sec.~\ref{sec3}, and derive
the modifications to the sum rules for $\lambda_{E,H}^2$.

\section{Nonperturbative corrections with dimension-6 operators}
\label{sec22}
At the leading accuracy in $\alpha_s$,
all the relevant power corrections to \eq{OPEgeneral}
with $X=F$
are generated 
from the 
second term in \eq{corfun0wick}, and 
the leading contribution in the ellipses in the middle line of \eq{corfun0wick2}
reads
\begin{equation}
\frac{1}{6}
x^\mu x^\nu x^\lambda \bra 0| \overline{q}(0) \Gamma_1 P_+ \Gamma_2 D_\mu D_\nu D_\lambda 
q(0) |0\ket ,
\label{corfun0wick30}
\end{equation}
which determines the power correction 
associated with the dimension-6 operator.
We exploit the following exact relation between matrix elements of the
local operators
($\Gamma$ is arbitrary gamma matrix)~\cite{grozin},
\bey
&& 
\!\!\!\!\!\!\!\!\!\!\!\!\!\!\!\!\!\!
\bra 0| \overline{q} \Gamma D_\mu D_\nu D_\lambda 
q |0\ket 
=\frac{ig^2}{576}
\bra 0|  \overline{q}   \gamma_\kappa T^a q  \sum_{q'}\overline{q}'\gamma^\kappa T^a q' |0\ket
\nonumber\\
&&\!\!\!\!\!\!\!\!\!
\times
\Tr\left[\Gamma \left(
g_{\mu\nu}\gamma_\lambda + g_{\nu \lambda} \gamma_\mu 
-5g_{\mu\lambda}\gamma_\nu 
-3i\varepsilon_{\mu \nu \lambda \rho}
\gamma^\rho \gamma_5
\right) \right] ,
\label{corfun0wick3}
\eey
which can be 
derived straightforwardly using the equations of motion,
$\Slash{D}q=0$, $D^\nu G_{\mu \nu}^a = g \sum_{q'}\overline{q}'\gamma_\mu T^a q'$,
where the summation $\sum_{q'}$ is over all quark flavors.
Thus, \eq{corfun0wick30} yields
\begin{equation}
\Tr\left[\Gamma_1\pp\Gamma_2\right]
\frac{ig^2}{1152}(v\cdot x)x^2
\bra 0| \overline{q}   \gamma_\kappa T^a q  \sum_{q'}\overline{q}'\gamma^\kappa T^a q' |0\ket ,
\label{corfun0wick4}
\end{equation}
which implies the new power-correction term in the RHS of Eq.~(\ref{OPEgeneral})
with $X=F$, given as~\cite{shuryak,neubert}
\begin{equation}
+ \frac{\pi C_F \alpha_s}{24 N_c \omega^4}\qcon^2 ,
\label{fsr0add0}
\end{equation}
where, as usual, the four-quark condensate in \eq{corfun0wick4}
is reduced to the square of $\qcon$ using the factorization approximation 
through the vacuum saturation.
As the result, 
the RHS of \eq{fsr0}
receives the new term~\cite{shuryak,neubert},
\begin{equation}
+\frac{\pi C_F \alpha_s}{72N_c M^3} \qcon^2 .
\label{fsr0add}
\end{equation}

\begin{figure}
\includegraphics[width=2.5cm,keepaspectratio]{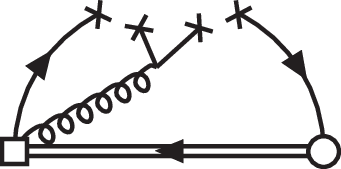}
\caption{A diagram for the OPE 
of the correlation function~(\ref{corfun}). At the leading accuracy in $\alpha_s$,
this diagram
determines the power correction 
induced by
the dimension-6 operators.}
\label{dim6}
\end{figure}

Similarly, we can calculate the new corrections
to the sum rules (\ref{treeSRformagnetic}) and (\ref{treeSRforsplitting})
for $\lambda_{E,H}^2$, induced by the dimension-6 operators:
at the leading accuracy in $\alpha_s$,
all the relevant power corrections to \eq{OPEgeneral}
with $X=3H,3S$ are generated from the second term in \eq{corfunwick}. 
The leading contribution in the ellipses in \eq{corfunwick2} 
is associated with the dimension-6 operators and reads
\bey
&&\!\!\!
x^\lambda \bra 0|\overline{q}(0) gG_{\mu\nu}(0) \Gamma_1 P_+ \Gamma_2 D_\lambda q(0)
|0\ket 
\nonumber\\
&&\!\!\!\!
= - \frac{iv\cdot x}{96}g^2 \bra 0| \overline{q}   \gamma_\kappa T^a q  \sum_{q'}\overline{q}'\gamma^\kappa T^a q' |0\ket
\left(
\Tr\left[\sig_{\mu\nu}\Gamma_1 \pp \Gamma_2 \right]\right.
\nonumber\\
&&\left. 
\;\;\;
+2
\Tr\left[(iv_\mu \gamma_\nu -iv_\nu \gamma_\mu)\Gamma_1 \pp \Gamma_2 \right]
\right) ,
\label{corfunwick3}
\eey
where the matrix element has been handled using 
\eq{corfun0wick3} with the indices $\mu$ and $\nu$
antisymmetrized.
Diagrammatically, this result is represented 
in Fig.\ref{dim6}.
\eq{corfunwick3} implies
the new power-correction term in the RHS of Eq.~(\ref{OPEgeneral}),
given as
\begin{equation}
+\frac{\pi C_F\alpha_s}{24N_c \omega^2}\qcon^2 ,
\;\;\;\;\;\;\;\;\;
+\frac{\pi C_F\alpha_s}{12N_c \omega^2}\qcon^2 ,
\label{dim6HS0}
\end{equation}
for $X=3H, 3S$, respectively, so that
we find that Eqs.~(\ref{treeSRformagnetic}), (\ref{treeSRforsplitting})
receive the new terms,
\begin{equation}
+\frac{\pi C_F\alpha_s}{2N_c M} \qcon^2,
\;\;\;\;\;\;\;\;\;
+\frac{\pi C_F\alpha_s}{N_c M} \qcon^2,
\label{dim6HS}
\end{equation} 
respectively, in their RHS.
\begin{figure}
\includegraphics[width=8.5cm,keepaspectratio]{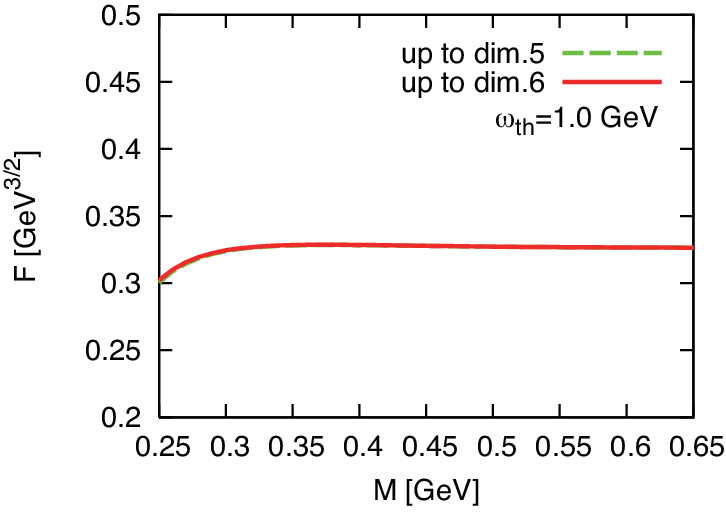}
\caption{Borel sum rule for $F$ using $\mu=1$~GeV, $\omega_{\rm th}=1.0\gev$, 
and $\bar\Lambda=0.55 \gev$, with and without adding \eq{fsr0add}
to the RHS of \eq{fsr0}. Two curves are almost indistinguishable.
}
\label{dim6-F}
\end{figure}

We now discuss the effect of 
the above-obtained power corrections due to the dimension-6 condensates
on the corresponding sum rules.
First of all, when including \eq{fsr0add} in the sum rule (\ref{fsr0}),
only this new term 
is of $O(\alpha_s)$,
in contrast to the other terms arising in the RHS of \eq{fsr0}, and, actually,
the effect of this new term
turns out to be completely negligible.
This is demonstrated in Fig.~\ref{dim6-F}; here,
the dashed curve is almost indistinguishable from the solid curve,
and the former is same as the solid curve in Fig.~\ref{ope-conv-F}.
\begin{figure}
\includegraphics[width=8.5cm,keepaspectratio]{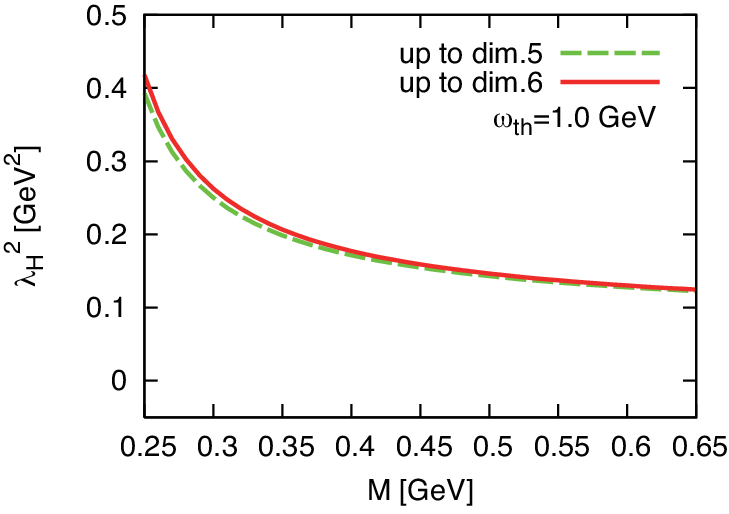}
\caption{
Borel sum rule for $\lambda_H^2$ with $\mu=1$~GeV and $\omega_{\rm th}=1.0\gev$,
calculating 
\eq{treeSRformagnetic} with and without taking into account \eq{dim6HS},
and 
dividing the results by \eq{fsr0}.
}
\label{dim6-H}
\end{figure}
\begin{figure}
\includegraphics[width=8.5cm,keepaspectratio]{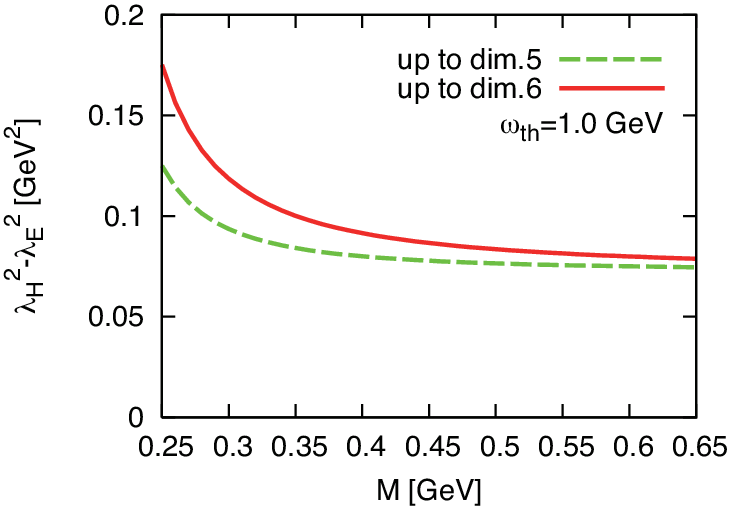}
\caption{Same as Fig.~\ref{dim6-H}, but for $\lambda_H^2-\lambda_E^2$,
calculating 
\eq{treeSRforsplitting} with and without taking into account \eq{dim6HS},
and 
dividing the results by \eq{fsr0}.
}
\label{dim6-S}
\end{figure}

For the case of the sum rules
(\ref{treeSRformagnetic}) and (\ref{treeSRforsplitting})
for $\lambda_{E,H}^2$, 
the terms in the RHS, except the term associated with the mixed condensate
$\mixqn$, are of 
$O(\alpha_s)$
similarly 
as the new contributions of \eq{dim6HS}.
The quantitative roles of these new terms are shown in Figs.~\ref{dim6-H},
\ref{dim6-S}:
we calculate \eq{treeSRformagnetic} with and without taking into account \eq{dim6HS},
and divide both the results by \eq{fsr0},
yielding the solid and dashed curves, respectively, plotted in Fig.~\ref{dim6-H}.
The similar calculation based on \eq{treeSRforsplitting}
yields the solid and dashed curves in Fig.~\ref{dim6-S}; note that
the latter curve is same as the dot-dashed curve in Fig.~\ref{ope-conv-S}
because the dimension-5 condensate does not contribute to \eq{treeSRforsplitting}.
The new contributions of \eq{dim6HS}
enhance the values of $\lambda_H^2$ and $\lambda_H^2-\lambda_E^2$,
giving rise to some additional dependence on the Borel parameter $M$,
but the effect is not so significant.
Indeed, the comparison of Figs.~\ref{dim6-H} and
\ref{dim6-S} with Figs.~\ref{ope-conv-H} and \ref{ope-conv-S}, respectively,  
indicates that 
the effects of \eq{dim6HS} due to the dimension-6 condensates
are smaller than the dominant effects from the lower-dimensional condensates,
such that the convergence of the OPE (\ref{OPEgeneral}) with $X=3S$ as well as
$X=3H$
may be suggested at this level of power corrections.

\section{One-loop Wilson coefficients for
the dimension-5 operators}
\label{sec3}
\begin{figure}
\includegraphics[width=8.5cm,keepaspectratio]{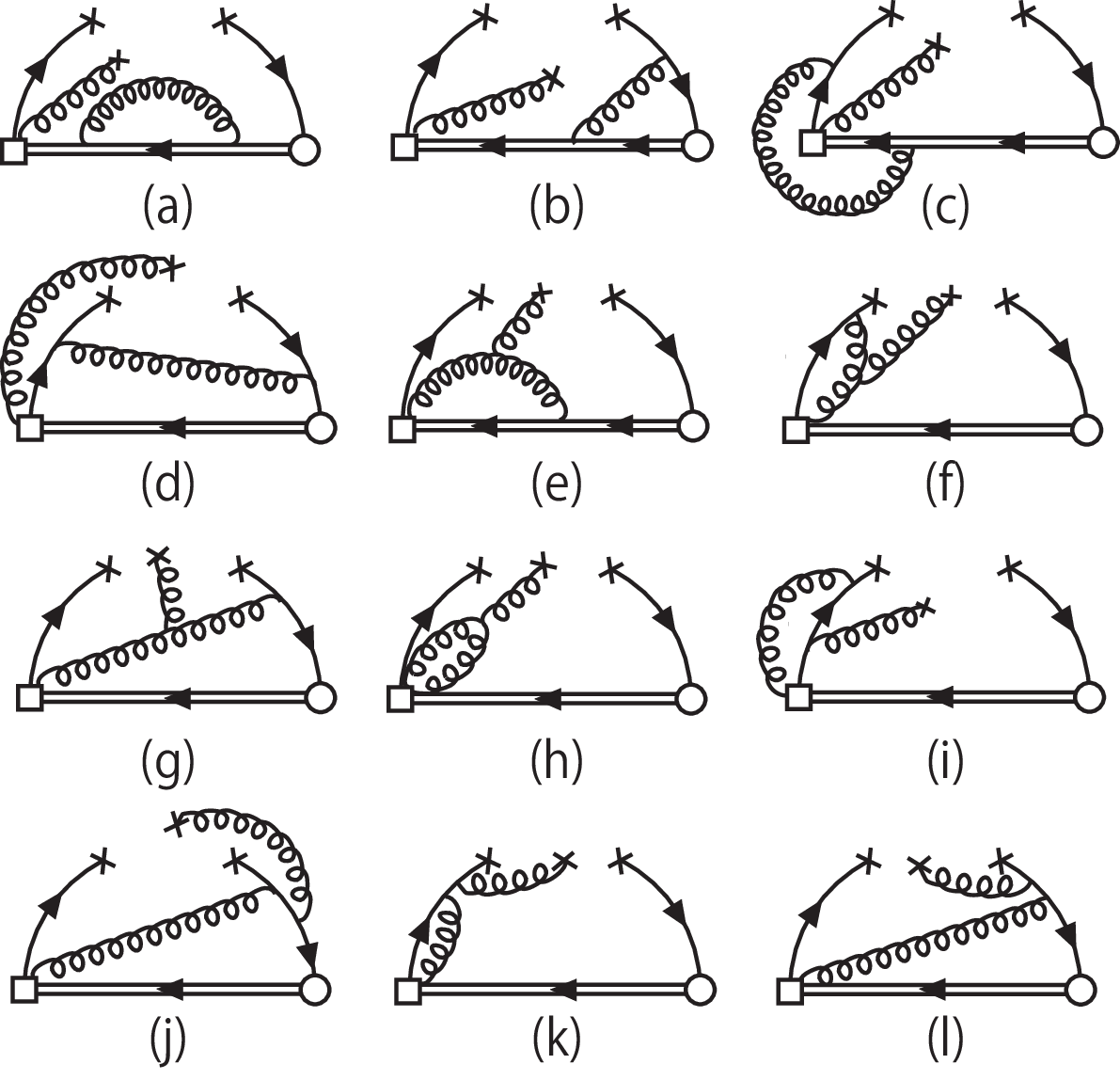}
\caption{Feynman diagrams for the one-loop matching
of the Wilson coefficients associated with the dimension-5 operators.} 
\label{correction-1}
\end{figure}
\begin{figure}
\includegraphics[width=8.5cm,keepaspectratio]{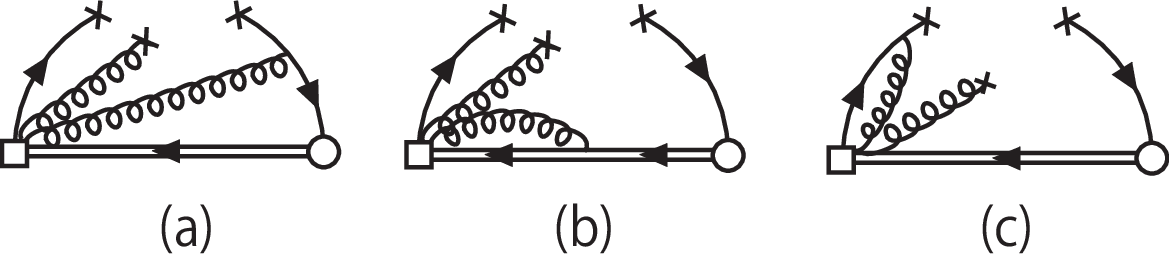}
\caption{Examples of the vanishing diagrams for the one-loop corrections to the correlator~(\ref{corfun}) 
in the Fock-Schwinger gauge.}
\label{correction-2}
\end{figure}

The Wilson coefficients arising in the OPE (\ref{OPEgeneral}) are in general
expressed as a power series in $\alpha_s$, 
\begin{equation}
C_k^{X}(\omega) = C_k^{X(0)}(\omega) + \frac{\alpha_s}{4\pi} C_k^{X(1)}(\omega)
+ \cdots .
\label{ps}
\end{equation} 
In particular, for the case with $X=3H, 3S$, the formulas (\ref{WC3H<5}), (\ref{WC3S<5}) and (\ref{dim6HS0})
indicate $C_k^{X(0)}(\omega)=0$,
for the coefficients associated with the operators of dimension $d\le6$, 
except for $k=\sigma$ and $X=3H$. Those formulas also give
the explicit nonzero results of
$C_k^{3H(1)}(\omega)$ and $C_k^{3S(1)}(\omega)$, 
except for the case with $k=\sigma$, whose results have been unknown.
In this section, 
we derive 
$C_\sigma^{3H(1)}(\omega)$ and $C_\sigma^{3S(1)}(\omega)$,
performing the one-loop matching calculation.
The OPE~(\ref{OPEgeneral}) with this result allows us to 
construct the sum rules for $\lambda_H^2$ and $\lambda_H^2-\lambda_E^2$,
taking into account all the relevant $O(\alpha_s)$ effects.

The Feynman diagrams in Fig.~\ref{correction-1} represent
the one-loop corrections to the correlation function in the LHS of \eq{corfun},
which are relevant to the matching to derive $C_\sigma^{X(1)}(\omega)$ for $X=3H, 3S$;
here, the corrections due to self-energy insertions into the quark or gluon external fields 
have been omitted, because the corresponding contributions eventually cancel in the matching.
We calculate those Feynman diagrams 
in $D=4+2\epsilon$ dimensions and derive $C_\sigma^{X(1)}(\omega)$
in the $\overline{\rm MS}$ scheme.
It is convenient to use the Fock-Schwinger
gauge for the classical background gluon field, as in Sec.~\ref{sec2}. 
Then, the diagrams containing the subdiagram (a) of Fig.~\ref{subdiagram}, as well as the subdiagram (b) or (c), 
vanish. Also, for the correlator~(\ref{corfun}),
the diagrams in Fig.~\ref{correction-2} vanish;
in Fig.~\ref{correction-2}, the vertex
arising 
from the quark-gluon current 
$\overline{q}(0)\Gamma_1 gG_{\mu\nu}(0) h_v(0)$ vanishes using 
\eq{fs}, and, indeed, the corresponding contribution is absent from 
the relevant propagator~(\ref{gprop}).

Here, for later use, we mention the contribution of the tree diagram~(d) in Fig.~\ref{treediagram}
to the correlator in the LHS of \eq{corfun}, for the case of $D$ 
dimensions and under 
the classical background fields
$q(x)$, $\bar{q}(x)$ and $G_{\mu\nu}(x)$ for quarks and gluons;
the corresponding contribution
reads
(see \eq{hprop}),
\bey
&& \!\!\!\!\!\!\!\!\!\!\!
i \int d^D x e^{-i\omega v\cdot x}
\theta(-v\cdot x)\delta^{(D)}\left(x_\perp\right)
\nonumber\\
&&\;\;\;\;\;\;\;\;\;\;\;\;\;\;\;\;\;\;\;\;\;\;\;\;
\times
\overline{q}(0) gG_{\mu\nu}(0) \Gamma_1 P_+ \Gamma_2 q (0)
\nonumber\\
&&\!\!\!\!\!
= - \frac{1}{2}\Tr\left[\sig_{\mu\nu}\Gamma_1 \pp \Gamma_2 \right] \widehat{\Pi}^{\rm tree}_{3H}(\omega) 
\nonumber\\
&&\!
-  \frac{1}{2}
\Tr\left[(iv_\mu \gamma_\nu -iv_\nu \gamma_\mu)\Gamma_1 \pp \Gamma_2 \right]
\widehat{\Pi}^{\rm tree}_{3S}(\omega),
\label{Phi}
\eey
where the extraction of the relevant scalar piece 
from the arising combination of the classical fields is implicit in the LHS,
yielding the RHS that is expressed in a similar form as in the RHS of \eq{corfun}, with
\bey
&& \widehat{\Pi}^{\rm tree}_{3H}(\omega) 
=\frac{1}{2D(D-1)\omega} 
{\cal O} ,
\nonumber\\
&& \widehat{\Pi}^{\rm tree}_{3S}(\omega) 
=0 ,
\label{pitree}
\eey
in terms of the dimension-5 quark-gluon scalar operator,
\begin{equation}
{\cal O}= \overline{q} gG \cdot \sigma q .
\label{calo}
\end{equation}
When $D\rightarrow 4$ and ${\cal O}\rightarrow \mixqn$, \eq{Phi} reduces to
the corresponding contribution using the first
term in the RHS of \eq{corfunwick2}.

Now, it is straightforward to calculate the diagrams in Fig.~\ref{correction-1}
combining the relevant propagators (\ref{hprop}), (\ref{qprop}), (\ref{gprop}) with other 
(familiar) building blocks, and it is convenient to 
perform 
the loop integrations in the coordinate space:
the diagrams~(e)-(g) and (i)-(j) in Fig.~\ref{correction-1} represent the contributions
generated by the second term in the RHS
of \eq{gprop} and of \eq{qprop}, respectively, and the diagram~(h) 
can be calculated similarly; on the other hand,
from the ``nonlocal quark condensate'' 
contributions 
contained in the ellipses in \eq{corfunwick},
the diagrams~(k), (l) are generated 
as the subleading terms
in the Taylor expansion
similar as in \eq{corfun0wick2}.
We note that the diagrams~(a)-(c) and (e) all give the UV-divergent results,
while the diagrams~(g), (j) and (l) give the IR-divergent results.
Each of the diagrams (f), (h), (i) and (k) vanishes as a result of the 
``canceling'' UV and IR poles, $1/\eps_{UV} -1/\eps_{IR}$,
arising from the scaleless loop integral.
On the other hand, the diagram~(d) yields the result of $O(\eps)$ and 
vanishes as $D \rightarrow 4$.
On the LHS of \eq{corfun}, 
in addition to the contributions from all the diagrams in Fig.~\ref{correction-1},
we have also the counter-term contribution, 
\bey
&&\!\!\!
\left[-\left(Z_h -1\right)
+\left( \frac{\sqrt{Z_2}\sqrt{Z_h}}{Z_J}-1 \right)
\right]
\nonumber\\
&&
\times\left( -\frac{1}{2} 
\Tr\left[\sig_{\mu\nu}\Gamma_1 \pp \Gamma_2 \right] \widehat{\Pi}^{\rm tree}_{3H}(\omega)\right)  + \cdots,
\label{ct}
\eey
with $\widehat{\Pi}^{\rm tree}_{3H}(\omega)$ defined as Eqs.~(\ref{Phi}),~(\ref{pitree}).
$Z_h$ and $Z_2$ are the quark-field renormalization constants
for heavy- and light-quarks, respectively, in the $\overline{\rm MS}$ scheme using
the Feynman gauge for the quantum part of the gluon field, as
\begin{equation}
Z_h =1-\frac{C_F \alpha_s}{2\pi \hat{\eps}}, 
\;\;\;\;\;\;\;\;\;\;\;\;
Z_2 =1+\frac{C_F \alpha_s}{4\pi \hat{\eps}}, 
\label{z2}
\end{equation}
at one-loop order~\cite{neubert0},
where
$1/\hat{\eps} \equiv 1/\eps +\gamma_E -\ln4\pi$, with $\gamma_E$
the Euler constant,
and, similarly, 
$Z_J$ is the renormalization constant 
for 
the heavy-light
current operator $J\equiv \overline{h}_v \Gamma_2 q$ in \eq{corfun},
as
\begin{equation}
J^{\rm ren}=\frac{1}{Z_J}J^{\rm bare}, 
\;\;\;\;\;\;\;\;\;\;
Z_J=1-\frac{3 C_F \alpha_s}{8\pi \hat{\eps}},
\label{zhl}
\end{equation}
connecting the renormalized and bare operators at one-loop order~\cite{neubert0}.
In \eq{ct}, the remaining counter terms,
associated with the renormalization of 
the quark-gluon three-body current operator $\overline{q} \Gamma_1 gG_{\mu\nu} h_v$
in \eq{corfun}, are represented by the ellipses
whose explicit formula is given in \eq{ctct} in Appendix~\ref{operatorrenorm}.

We combine the sum of the contributions of all the diagrams in Fig.~\ref{correction-1}
with the counter-term contribution~(\ref{ct}),
and decompose the result as in the RHS of \eq{corfun}, denoting
the corresponding two correlation functions
as $\widehat{\Pi}^{\mbox{\scriptsize 1-loop}}_{3H}(\omega)$
and $\widehat{\Pi}^{\mbox{\scriptsize 1-loop}}_{3S}(\omega)$
in place of $\Pi_{3H,3S}(\omega)$.
We obtain,
for $\omega <0$,
\bey
 &&\!\!\!\!\!\!\!\!\!\!\! 
\widehat{\Pi}^{\mbox{\scriptsize 1-loop}}_{3H}(\omega)= 
\frac{\alpha_s}{96\pi\omega} 
\left[
\frac{2}{N_c\hat{\eps}}
-\left( N_c
- \frac{6}{N_c}\right)
\ln\frac{-2\omega}{\mu}
\right.
\nonumber\\
&& \;\;\;\;\;\;\;\;\;\;\;\;
\left.
+ \frac{5}{2}N_c- \frac{10}{3N_c}+\cdots 
\right]{\cal O} ,
\label{UVfinitePi1}\\
&& \!\!\!\!\!\!\!\!\!\! 
\widehat{\Pi}^{\mbox{\scriptsize 1-loop}}_{3S}(\omega)=
\frac{\alpha_s}{96\pi \omega} 
\left[
N_c\ln\frac{-2\omega}{\mu}
+ \frac{1}{2N_c}+\cdots\right]{\cal O} ,
\label{UVfinitePi2}
\eey
with the ellipses 
denoting the terms that
vanish as $\eps \rightarrow 0$.
Here and below, 
$\mu$ is the $\overline{\rm MS}$ scale
and 
$\alpha_s\equiv\alpha_s(\mu)$. 
In the calculation to derive these results,
we observe that the UV poles from the diagrams~(a) and (b) 
are canceled, respectively, by the first and second counter-terms
in \eq{ct}. Similarly,
the UV poles from the diagrams~(c) and (e),
as well as the UV poles from the diagrams~(f), (h), (i) and (k)
due to the above-mentioned $1/\eps_{UV} -1/\eps_{IR}$ structure,
are completely canceled by the UV poles arising in the ellipses
in \eq{ct}, i.e., by the counter terms~(\ref{ctct}).
Then, 
the remaining $1/\eps$ poles from the diagrams in Fig.~\ref{correction-1}
are the IR poles only, as it should be,
and the sum of all those IR poles yields the
$1/\eps$ pole in the above results~(\ref{UVfinitePi1}), (\ref{UVfinitePi2}).

As a useful cross-check of our results~(\ref{UVfinitePi1}), (\ref{UVfinitePi2}),
we performed the corresponding NLO calculation
also for the correlation function,
\begin{equation}
i\int d^D x e^{i\omega v\cdot x}
\bra 0|T\left[\overline{q}(x)\Gamma_1 gG_{\mu\nu}(x) h_v(x)\ \h_v(0) \Gamma_2 q(0)
\right]
|0\ket .
\label{corfuncheck}
\end{equation}
Because the Fock-Schwinger gauge, $x^\mu A_\mu(x)=0$, used in the present paper,
violates translational invariance,
the calculation of \eq{corfuncheck}
does not coincide on a diagram-by-diagram basis with that
of the LHS in \eq{corfun};
e.g., for the correlator~(\ref{corfuncheck}), the contributions of the diagrams~(a)-(c) 
in Fig.~\ref{correction-2} 
do not vanish.
With the similar technique as above and with $\eps_{IR}=\eps_{UV}$, 
we calculate the contributions to \eq{corfuncheck} 
taking into account all the relevant diagrams, i.e.,
the diagrams in Figs.~\ref{correction-1},~\ref{correction-2}, and some other 
nonvanishing diagrams.
We find that this calculation yields the results
identical to 
Eqs.~(\ref{UVfinitePi1}), (\ref{UVfinitePi2}).
Indeed, for example, considering 
the diagram~(e) in Fig.~\ref{correction-1} and the diagram~(b) in Fig.~\ref{correction-2},
the sum of the contributions of these diagrams for \eq{corfuncheck} coincides with the corresponding sum for \eq{corfun}.
This reflects the fact that the translational invariance is restored 
in a gauge-invariant subset of the diagrams;
note that the contributions of the diagrams involving the subdiagrams in Fig.~\ref{subdiagram}
vanish for \eq{corfuncheck}, as well as for \eq{corfun}.
Similarly, the diagrams~(g), (j), (l) in Fig.~\ref{correction-1}, the diagram~(a) in Fig.~\ref{correction-2}, and some other diagrams
form a gauge-invariant subset,
such that they are obtained from the diagram (b) in Fig.~\ref{treediagram} by attaching an external gluon line in all possible ways,
and we observe that the sum of the contributions of those diagrams is identical between \eq{corfuncheck} and \eq{corfun}.
With $\eps_{IR}=\eps_{UV}$, 
each of the remaining diagrams in Figs.~\ref{correction-1},~\ref{correction-2}
actually yields the identical result for both Eqs.~(\ref{corfuncheck}) and~(\ref{corfun}).

The matching relations of our above results (\ref{Phi})-(\ref{pitree}),~(\ref{UVfinitePi1}) and (\ref{UVfinitePi2})
with the corresponding term in the OPE~(\ref{OPEgeneral}) read ($X=3H, 3S$),
\bey
\widehat{\Pi}^{\mbox{\scriptsize tree}}_{X}(\omega)+
\widehat{\Pi}^{\mbox{\scriptsize 1-loop}}_{X}(\omega)
=C_\sigma^X(\omega) {\cal O}^{\rm ren} ,
\label{pert}
\eey
for $-\omega \gg \Lambda_{\rm QCD}$,
with the Wilson coefficients expressed as \eq{ps},
and the renormalized composite operator ${\cal O}^{\rm ren}$ corresponding to \eq{calo},
such that $\bra 0| {\cal O}^{\rm ren}|0\ket = \mixqn$. 
Here, the renormalized operator ${\cal O}^{\rm ren}$ is related to the bare operator ${\cal O}^{\rm bare}$,
and to the operator ${\cal O}$ which arises in Eqs.~(\ref{pitree}),~(\ref{UVfinitePi1}) and (\ref{UVfinitePi2})
and is composed of the classical (renormalized) constituent fields, as 
\begin{equation}
{\cal O}^{\rm ren}= \frac{1}{Z_{\cal O}}{\cal O}^{\rm bare} = \frac{Z_2}{Z_{\cal O}}{\cal O},
\label{caloren}
\end{equation}
with $Z_2$ of \eq{z2}, noting that the combination $gG_{\mu \nu}$ 
is not renormalized in the background field method~\cite{abott}.
To obtain $Z_{\cal O}$, we performed the one-loop renormalization of the dimension-5 quark-gluon-mixed operator~(\ref{calo}) in
the present framework with the $\overline{\rm MS}$ scheme, and the result is
(see the discussion below \eq{mm} in Appendix~\ref{operatorrenorm})
\begin{equation}
Z_{\cal O}=1+ \frac{\alpha_s}{8\pi \hat{\eps}}\left(N_c -\frac{5}{N_c} \right) ,
\label{zo}
\end{equation}
which coincides with the corresponding result calculated in Ref.~\cite{narison}.
Now, the matching of the $O(\alpha_s^0)$ terms of both sides in \eq{pert}
immediately yields
\bey
&&C_\sigma^{3H(0)}(\omega)=\frac{1}{4 (2+ \epsilon) (3+ 2 \epsilon) \omega}, 
\nonumber\\
&&C_\sigma^{3S(0)}(\omega) =0, 
\label{csigma0}
\eey
for arbitrary $\eps$, and
these results reproduce the corresponding formulas 
in Eqs.~(\ref{WC3H<5}) and~(\ref{WC3S<5}) as $\eps \rightarrow 0$.
The one-loop matching due to
the $O(\alpha_s)$ terms in \eq{pert} leads to the relation,
\begin{equation}
\frac{\alpha_s}{4\pi} C_\sigma^{3X(1)}(\omega) 
=\frac{\partial \widehat{\Pi}^{\mbox{\scriptsize 1-loop}}_{3X}(\omega)}{\partial {\cal O}}
- \left(\frac{Z_2}{Z_{\cal O}}-1\right) C_\sigma^{3X(0)}(\omega) ,
\label{matching1}
\end{equation}
the both sides of which are finite as $\eps \rightarrow 0$: 
for $X=3H$, the second term in the RHS serves to cancel 
the $1/\eps$ pole arising in the first term (see \eq{UVfinitePi1}),
while, for $X=3S$, \eq{csigma0} implies that
$C_\sigma^{3S(1)}(\omega)$ is directly given by the coefficient of ${\cal O}$
in \eq{UVfinitePi2}.
Substituting the $\eps \rightarrow 0$ limit of Eqs.~(\ref{csigma0})
and (\ref{matching1}) into \eq{ps}, we obtain the final form of 
the corresponding NLO Wilson coefficients,
\bey
C_\sigma^{3H}(\omega)&=& \frac{1}{24\omega}\left[1-\frac{\alpha_s}{4\pi} 
\left\{\left(N_c-\frac{6}{N_c}\right) \ln\frac{-2\omega}{\mu}
\right.\right.
\cr&& \;\;\;\;\;\;\;\;\;\;\;\;
\left.\left.
-\frac{5}{2}N_c+\frac{1}{N_c}\right\} \right] ,
\label{finalc3H}\\
C_\sigma^{3S}(\omega)&=&\frac{\alpha_s}{96\pi\omega}
\left[N_c \ln\frac{-2\omega}{\mu}
+ \frac{1}{2N_c}\right]
\label{finalc3S},
\eey
in the $\overline{\rm MS}$ scheme.

\section{Renormalization-group improvement and 
Borel analysis}
\label{sec5}

We substitute the new results~(\ref{finalc3H}) and (\ref{finalc3S})
into the Wilson coefficient $C_\sigma^X(\omega)$ 
of \eq{OPEgeneral}, while, for the other coefficients
$C_I^X(\omega)$, $C_q^X(\omega)$, and $C_G^X(\omega)$,
we have the 
corresponding formulas in 
Eqs.~(\ref{WC3H<5}) and (\ref{WC3S<5}); furthermore, we add the new power-correction term with (\ref{dim6HS0})
to the RHS of \eq{OPEgeneral}.
The result gives our upgraded OPEs for the correlation functions in \eq{corfun}, 
taking into account the operators of dimension $d \leq 6$ and the associated Wilson coefficients
to the $O(\alpha_s)$ accuracy.
Now, we use these new results for the OPE to derive the sum rules for $\lambda_H^2$ and $\lambda_H^2-\lambda_E^2$:
we substitute these OPEs
into Eqs.~(\ref{BorelSR1}) and (\ref{BorelSR2}); here,
it is straightforward to calculate the discontinuities of the coefficients~(\ref{finalc3H}) and (\ref{finalc3S})
across the cut along the line $\omega>0$ in the complex $\omega$ plane,
reexpressing the logarithmic contributions as $(2/\omega) \ln(-2\omega/\mu)= (d/ d\omega)\ln^2(-2\omega/\mu)$.
As a result, we obtain  
the new formulas of the Borel sum rules for $\lambda_H^2$ and $\lambda_H^2-\lambda_E^2$,
\bey
F^2(\mu)\lambda_H^2(\mu) e^{-\bar\Lambda/M}&&
\!\!\!\!\!\!
=
-\frac{2\alpha_s}{\pi^3}
N_cC_FM^5W^{(4)}(\frac{\omega_{\rm th}}{M})
\nonumber\\
&&\!\!\!\!\!\!\!\!\!
- \frac{3\alpha_s}{\pi}C_FM^2\qcon W^{(1)}(\frac{\omega_{\rm th}}{M}) 
\nonumber\\
&& \!\!\!\!\!\!\!\!\!\!\!\!\!\!\!\!\!
+\frac{M}{4} 
\gc W^{(0)}(\frac{\omega_{\rm th}}{M}) 
\nonumber\\ &&
\!\!\!\!\!\!\!\!\!\!\!\!\!\!\!\!\!\!\!\!\!\!\!\!\!\!\!\!\!\!\!\!
\left.-\frac{1}{4} \mixqn \right[1
-\frac{\alpha_s}{4\pi}
\left\{-\frac{5}{2}N_c+\frac{1}{N_c} \right.
\nonumber\\ &&\!\!\!\!\!\!\!\!\!\!\!\!\!\!\!\!\!\!\!\!\!\!\!\!\!\!\!\!\!\!\!\!
\left.\left.+\left(N_c-\frac{6}{N_c}\right)
\left(\ln\frac{2M}{\mu e^{\gamma_E}}
-\Gamma(0,\frac{\omega_{\rm th}}{M})\right)
\right\}\right]
\nonumber\\ &&
\!\!\!\!\!\!\!\!\!\!\!\!\!\!\!\!\!\!\!\!\!\!\!\!\!
+\frac{\pi C_F\alpha_s}{2N_c M} \qcon^2,
\label{SRformagnetic}
\eey
and
\bey
\lefteqn{F^2(\mu)\left[\lambda_H^2(\mu)-\lambda_E^2(\mu)\right]e^{-\bar\Lambda/M}}
\nonumber\\
&& \;\;\;\;\;\;\;\;\;\;\;\;
=\frac{2\alpha_s}{\pi^3}
N_cC_FM^5W^{(4)}(\frac{\omega_{\rm th}}{M})
\nonumber\\
&& \;\;\;\;\;\;\;\;\;\;\;\;
-\frac{3\alpha_s}{\pi}C_FM^2 \qcon W^{(1)}(\frac{\omega_{\rm th}}{M})
\nonumber\\
&& \;\;\;\;\;\;\;\;\;\;\;\;
+ \frac{M}{4} \gc W^{(0)}(\frac{\omega_{\rm th}}{M}) 
\nonumber\\ &&
\!\!
-\mixqn \frac{\alpha_s}{16\pi}
\left[N_c\left(\ln\frac{2M}{\mu e^{\gamma_E}}
-\Gamma(0,\frac{\omega_{\rm th}}{M})\right)
+\frac{1}{2N_c}\right]
\nonumber\\ &&
\;\;\;\;
+\frac{\pi C_F\alpha_s}{N_c M} \qcon^2,
\label{SRforsplitting}
\eey
where $W^{(m)}(\omega_{\rm th}/M)$ is defined as \eq{Wm} and 
\begin{equation}
\Gamma(a,z)=\int^\infty_z dt \,t^{a-1}e^{-t}
\end{equation}
is the incomplete Gamma function.
These two sum rules, (\ref{SRformagnetic}) and (\ref{SRforsplitting}),
are the new results that 
take into account the operators of dimension $d \leq 6$ and the associated Wilson coefficients
to the $O(\alpha_s)$ accuracy: 
compared with the previous results~(\ref{treeSRformagnetic}) and (\ref{treeSRforsplitting})
that correspond to Grozin-Neubert's sum rule formulas~\cite{GN},
Eqs.~(\ref{SRformagnetic}) and~(\ref{SRforsplitting}) receive the $O(\alpha_s)$ corrections
associated with the dimension-5 quark-gluon-mixed condensate $\mixqn$, as well as \eq{dim6HS}
due to the 
dimension-6 four-quark condensate $\qcon^2$.
In particular, the former corrections bring an explicit 
dependence on the scale $\mu$ to the RHS of Eqs.~(\ref{SRformagnetic}) and~(\ref{SRforsplitting}), 
through the logarithmic term, $\ln(2M/\mu e^{\gamma_E})$:
one can show
that $\lambda_{E,H}^2(\mu)$ determined by our formulas (\ref{SRformagnetic}) and~(\ref{SRforsplitting}) 
satisfy the renormalization-group equations of Eqs.~(\ref{elam}),~(\ref{mm}),
taking into account the derivative of 
the above logarithm $\ln(2M/\mu e^{\gamma_E})$,
as well as the scale dependence of the other terms controlled by 
the nontrivial anomalous dimensions:
\bey
&&\left(\mu\frac{d}{d\mu}+ \gamma_{J} (\alpha_s)\right)
F(\mu)= 0 ,
\label{evol:F}\\
&&\left(\mu\frac{d}{d\mu}+ \gamma_{q}(\alpha_s)\right)\qcon(\mu)=0 ,
\label{evol:qcon}\\
&&\left( \mu\frac{d}{d\mu}+\gamma_{\sigma} (\alpha_s)
 \right) \mixqn(\mu)=0 ,
\label{evol:mixq}\\
&&
\!\!\!\! \!\!\!\! \!\!\!\!\! \!\!\!\! 
\gamma_{k} (\alpha_s)= \gamma_{k0}\frac{\alpha_s}{4\pi}
+\gamma_{k1}\left(\frac{\alpha_s}{4\pi}\right)^2\ + \cdots
\;\;\;\;\; (k=J, q, \sigma),
\label{ad}
\eey
where
\bey
&&\gamma_{J0} =-3C_F ,
\;\;\;
\gamma_{q0}=-  6C_F , \;\;\;
\gamma_{\sigma 0}=N_c -\frac{5}{N_c} ,
\label{J0sigma0}\\
&&
\!\!\!\! \!\!\!\! 
\gamma_{J1}=C_F\left[\left( \frac{5}{2}-\frac{8}{3}\pi^2 \right)C_F\right.
\nonumber\\&&
\left. \;\;\;\;\;\;\;\;\;\;\;\; +
\left( \frac{2}{3}\pi^2  -\frac{49}{6} \right)N_c+\frac{5}{3}N_f\right],
\label{J1}\\
&&
\!\!\!\! \!\!\!\! 
\gamma_{q1}=- C_F\left(3 C_F+\frac{97}{3} N_c-\frac{10}{3} N_f\right) ,
\label{q1}
\eey
while $\gamma_{\sigma 1}$ is not available.
For the present purpose to confirm \eq{elam} with the one-loop mixing matrix~(\ref{mm}),
the explicit forms of $\gamma_{J0}$ and $\gamma_{\sigma 0}$ in \eq{J0sigma0},
combined with $d\bar{\Lambda}/d\mu=0$~\cite{neubert0}, are sufficient,
where the former 
are immediate consequences of the corresponding renormalization constants (\ref{zhl}), (\ref{zo})~\footnote{$\gamma_{\sigma 0}$
has been obtained in Ref.~\cite{narison} and, independently and simultaneously, in Ref.~\cite{morozov}. $\gamma_{J1}$ is obtained 
in Ref.~\cite{gammaj1}. See also Refs.~\cite{neubert0,ElKhadra}.}.
Apparently, Grozin-Neubert's sum-rule formulas given by 
Eqs.~(\ref{treeSRformagnetic}) and (\ref{treeSRforsplitting})
do not obey such renormalization-group  
property.
Thus, our sum rules (\ref{SRformagnetic}) and~(\ref{SRforsplitting}) allow the first nonperturbative 
estimate of the HQET parameters 
$\lambda_{E,H}^2(\mu)$
with the correct $\mu$-dependence implemented.
We emphasize that the new $O(\alpha_s)$ corrections in Eqs.~(\ref{SRformagnetic}) and~(\ref{SRforsplitting}),
associated with the dimension-5 quark-gluon-mixed condensate $\mixqn$,
play essential roles to reproduce renormalization-group equations of Eqs.~(\ref{elam}),~(\ref{mm})
in the QCD sum-rule framework.

It is worth comparing this remarkable property of Eqs.~(\ref{SRformagnetic}) and~(\ref{SRforsplitting}) 
with the situation for the case of the sum rules of the decay constant $F(\mu)$,
based on the correlator~(\ref{corfun0}):
$F(\mu)$ determined from the sum rule~(\ref{fsr0}), which was presented in Sec.~\ref{sec2} at the leading accuracy in $\alpha_s$,
does not obey \eq{evol:F},
even though we take into account 
the scale dependence of Eqs.~(\ref{evol:qcon}), (\ref{evol:mixq}) in \eq{fsr0}.
Now, including the higher-order contributions, the $O(\alpha_s)$ corrections 
for the relevant Wilson coefficients and
the dimension-6 condensate contribution~(\ref{fsr0add}),
\eq{fsr0} is modified into~\cite{neubert,bagan,broad}
\bey
F^2(\mu) e^{-\bar\Lambda/M}\!\!\!\!\!\!
&&=\frac{N_cM^3}{\pi^2}\int_0^{\omega_{\rm th}/M}dz z^2 e^{-z}
\cr&&
\times
\left[1+\frac{3C_F\alpha_s}{2\pi}\left(\ln\frac{\mu}{2Mz}+\frac{17}{6}+\frac{2\pi^2}{9}
\right)\right]
\cr&&
-\qcon\left[1+\frac{3C_F\alpha_s}{2\pi} 
\right]
\cr&&
+\frac{1}{16M^2}\mixqn
\cr&&
+\frac{\pi C_F \alpha_s}{72N_c M^3} \qcon^2 .
\label{fsr}
\eey
Here, in particular, the $O(\alpha_s)$ corrections arising in the second line
bring an explicit
dependence on the scale $\mu$, through the logarithm $\ln(\mu/ 2Mz)$.
Taking into account this new $\mu$-dependence, $F(\mu)$ determined by \eq{fsr}
obeys the renormalization-group equation~(\ref{evol:F}),
up to the corrections of $O(\alpha_s^2)$
and the small contributions from the condensates of dimension-5 and higher
(see Figs.~\ref{ope-conv-F},~\ref{dim6-F}).

The correct renormalization-group properties obeyed by
Eqs.~(\ref{SRformagnetic}), (\ref{SRforsplitting}) and (\ref{fsr})
allow us to improve  
these sum rules further,
such that the logarithmic effects 
associated with $\alpha_s \ln(M/\mu)$ are resummed to all orders.
This renormalization-group improvement is formally achieved by setting $\mu=\mu'$ ($\mu' \sim M$)
in Eqs.~(\ref{SRformagnetic}), (\ref{SRforsplitting}) and (\ref{fsr}),
followed by evolving the resulting HQET parameters and the condensates at the scale $\mu'$
to those at 
$\mu\sim 1$~GeV:
%
using the first two coefficients of the $\beta$ function,
\bey
&&\beta_0
=\frac{11}{3}N_c-\frac{2}{3}N_f ,
\nonumber\\
&&\beta_1=\frac{34}{3}N_c^2-\frac{10}{3}N_cN_f-2C_FN_f ,
\eey
with $N_f$ being the number of active flavors,
the corresponding renormalization-group-improved sum rule for the decay constant reads
\bey
&&
\!\!\!\!\!\!\!\!\!
F^2(\mu') 
e^{-\bar\Lambda/M}
\nonumber\\
&&
\!\!\!\!\!\!
=
\frac{N_cM^3}{\pi^2}\int_0^{\omega_{\rm th}/M}dz z^2 e^{-z}
\cr&&
\times
\left[1+\frac{3C_F\alpha_s(\mu')}{2\pi}\left(\ln\frac{\mu'}{2Mz}+\frac{17}{6}+\frac{2\pi^2}{9}
\right)\right]
\cr&&
\!\!\!\!\!\!
-\left( {\frac{{\alpha _s (\mu' )}}
{{\alpha _s (\mu )}}} \right)^{\frac{{\gamma_{q0} }}
{{2\beta _0 }}} \left[ {1 + \frac{{\alpha _s (\mu' ) - \alpha _s (\mu )}}
{{4\pi }}\frac{{\gamma _{q0} }}
{{2\beta _0 }}\left( {\frac{{\gamma _{q1} }}
{{\gamma _{q0} }} - \frac{{\beta _1 }}
{{\beta _0 }}} \right)} \right]
\nonumber\\
&&\;\;\;\;\;\;\;\;\;\;\;\;\;\;\;\;\;\;\;\;\;\;\;\;\;\;\;\;\;\;\;\;\;\;\;\;\;\;\;\;
\times\left\langle {\bar qq} \right\rangle (\mu )
\left[1+\frac{3C_F\alpha_s(\mu')}{2\pi} 
\right]
\cr&&
+\frac{1}{16M^2} \left( {\frac{{\alpha _s (\mu' )}}
{{\alpha _s (\mu )}}} \right)^{\frac{{\gamma_{\sigma 0} }}
{{2\beta _0 }}}\mixqn (\mu )
\cr&&
+\frac{\pi C_F \alpha_s(\mu)}{72N_c M^3} \qcon^2(\mu) ,
\label{fsrRG}
\eey
to be combined with 
\bey
&&
\!\!\!\!\!\!\!\!\!
F^2(\mu)=F^2(\mu') \left( {\frac{{\alpha _s (\mu )}}
{{\alpha _s (\mu')}}} \right)^{\frac{\gamma_{J0}}{\beta_0}} 
\biggr[ 1+  \frac{{\alpha _s (\mu ) - \alpha _s (\mu' )}}
{{4\pi }}
\nonumber\\
&&\left. \;\;\;\;\;\;\;\;\;\;\;\;\;\;\;\;\;\;\;\;\;\;\;\;\;\;\;\;\;\;\;\;\;\;
\times \frac{{\gamma_{J0}}}
{{\beta_0 }}\left( {\frac{{\gamma_{J1}}}
{{\gamma_{J0}}} - \frac{{\beta_1 }}
{{\beta_0 }}} \right) \right] .
\label{evolF2}
\eey
Here, we have neglected the unknown NLO-level effects 
associated with the dimension-5 quark-gluon-mixed condensate, i.e.,
the corresponding one-loop coefficient function and 
two-loop anomalous dimension,
and also neglected the running of the 
dimension-6 four-quark condensate.
Up to these small effects,
\eq{fsrRG} combined with \eq{evolF2} sums up the leading and next-to-leading
logarithms of the form 
$\alpha_s^n \ln^n(M/\mu)$ and $\alpha_s^{n+1} \ln^n(M/\mu)$.

As a result of the similar renormalization-group improvement for
the two sum rules (\ref{SRformagnetic}) and (\ref{SRforsplitting}),
we obtain
\bey
F^2(\mu')\lambda_H^2(\mu') e^{-\bar\Lambda/M}&&
\!\!\!\!\!\!
=
-\frac{2\alpha_s(\mu')}{\pi^3}
N_cC_FM^5W^{(4)}(\frac{\omega_{\rm th}}{M})
\nonumber\\
&&
\!\!\!\!\!\!\!\!\!\!\!\!\!\!\!\!\!\!\!\!\!\!\!\!
- \frac{3\alpha_s(\mu')}{\pi}C_FM^2\left( {\frac{{\alpha _s (\mu' )}}
{{\alpha _s (\mu )}}} \right)^{\frac{{\gamma_{q0} }}
{{2\beta _0 }}} \qcon(\mu) 
\nonumber\\
&& \!\!\!\!\!\!\!\!\!\!\!\!\!\!\!\!\!
\times W^{(1)}(\frac{\omega_{\rm th}}{M}) +\frac{M}{4} 
\gc W^{(0)}(\frac{\omega_{\rm th}}{M}) 
\nonumber\\ &&
\!\!\!\!\!\!\!\!\!\!\!\!\!\!\!\!
-\frac{1}{4} \left( {\frac{{\alpha _s (\mu' )}}
{{\alpha _s (\mu )}}} \right)^{\frac{{\gamma_{\sigma 0} }}
{{2\beta _0 }}}\mixqn (\mu )
\nonumber\\
&&
\!\!\!\!\!\!\!\!\!\!\!\!\!\!\!\!
\times\left[1
-\frac{\alpha_s(\mu')}{4\pi}
\left\{-\frac{5}{2}N_c+\frac{1}{N_c} \right.\right.
\nonumber\\ &&\!\!\!\!\!\!\!\!\!\!\!\!\!\!\!\!\!\!\!\!\!\!\!\!\!\!\!\!\!\!\!\!
\left.\left.+\left(N_c-\frac{6}{N_c}\right)
\left(\ln\frac{2M}{\mu' e^{\gamma_E}}
-\Gamma(0,\frac{\omega_{\rm th}}{M})\right)
\right\}\right]
\nonumber\\ &&
\!\!\!\!\!\!\!\!\!\!\!\!\!\!\!\!\!\!\!\!\!\!\!\!\!
+\frac{\pi C_F\alpha_s(\mu)}{2N_c M} \qcon^2(\mu) ,
\label{RGformagnetic}
\eey
\bey
\lefteqn{F^2(\mu')\left[\lambda_H^2(\mu')-\lambda_E^2(\mu')\right]e^{-\bar\Lambda/M}}
\nonumber\\
&& \;\;\;\;\;\;\;\;\;\;\;\;
=\frac{2\alpha_s(\mu')}{\pi^3}
N_cC_FM^5W^{(4)}(\frac{\omega_{\rm th}}{M})
\nonumber\\
&& \;\;\;\;\;\;\;\;\;\;\;\;
-\frac{3\alpha_s(\mu')}{\pi}C_FM^2 \left( {\frac{{\alpha _s (\mu' )}}
{{\alpha _s (\mu )}}} \right)^{\frac{{\gamma_{q0} }}
{{2\beta _0 }}} \qcon (\mu) 
\nonumber\\
&&\;\;\;\;\;\;\;\;\;\;\;\;\;\;\;\;\;\;\;\;\;
\times W^{(1)}(\frac{\omega_{\rm th}}{M})
+ \frac{M}{4} \gc W^{(0)}(\frac{\omega_{\rm th}}{M}) 
\nonumber\\ &&
\;\;\;\;\;\;\;\;\;\;\;\;\;\;\;
-  \left( {\frac{{\alpha _s (\mu' )}}
{{\alpha _s (\mu )}}} \right)^{\frac{{\gamma_{\sigma 0} }}
{{2\beta _0 }}}\mixqn (\mu )
\nonumber\\
&&\;\;\;\;\;\;\;\;\;\;\;\;
\times
\frac{\alpha_s(\mu')}{16\pi}
\left[N_c\left(\ln\frac{2M}{\mu' e^{\gamma_E}}
-\Gamma(0,\frac{\omega_{\rm th}}{M})\right)
+\frac{1}{2N_c}\right]
\nonumber\\ &&
\;\;\;\;\;\;\;\;\;\;\;\;\;\;\;\;
+\frac{\pi C_F\alpha_s(\mu)}{N_c M} \qcon^2(\mu),
\label{RGforsplitting}
\eey
which are to be combined with
\begin{equation}
\left( 
\begin{array}{c}
\lambda_E^2(\mu)\\
\lambda_H^2(\mu)
\end{array}
\right) =
\left( {\frac{{\alpha _s (\mu )}}
{{\alpha _s (\mu' )}}} \right)^{\frac{\hat{\gamma}_0}
{{2\beta _0 }}}
\left( 
\begin{array}{c}
\lambda_E^2(\mu')\\
\lambda_H^2(\mu')
\end{array}
\right) .
\label{evolution}
\end{equation}
Here, the anomalous dimensions for $\lambda_{E,H}^2(\mu)$ are known only at one-loop as \eq{mm}.
We can divide Eqs.~(\ref{RGformagnetic}) and (\ref{RGforsplitting}) by \eq{fsrRG}
so as to eliminate the factors $F^2(\mu')e^{-\bar\Lambda/M}$,
before combining with the evolution~(\ref{evolution}).

We evaluate the renormalization-group-improved Borel sum rules
Eqs.~(\ref{fsrRG}), (\ref{RGformagnetic}) and (\ref{RGforsplitting}) 
with $\mu'=2M$ and $\mu=1$~GeV;
here, the choice $\mu'=2M$ allows us to sum up 
the relevant logarithmic contributions.
In addition to this default choice,
we also study the scale dependence of the results when varying $\mu'$ in a range 
with $\mu' \sim M$;
this gives us a rough estimate of the uncertainty of the results
due to the neglected part of order $\alpha_s^2$, 
and to the contributions of higher order.
We use the input values for the vacuum condensates 
as given in Table~\ref{parameter} in Sec.~\ref{sec2},
and use the two-loop expression for the running coupling 
$\alpha_s(\mu)$ with $\Lambda_{\rm QCD}^{(4)}=0.31$~GeV, so
that $\alpha_s(1~{\rm GeV})\simeq 0.47$, and $\alpha_s(m_B)\simeq 0.21$.

First of all, we note that 
\bey
&&\hat{F}\equiv F(\mu') \alpha _s (\mu')^{-\frac{\gamma_{J0}}{2\beta_0}} 
\left( 1 - \frac{\alpha _s (\mu' )}
{8\pi}\delta\right),
\label{rginvariant}\\
&&
\delta= \frac{\gamma_{J0}}{\beta_0}
\left( {\frac{{\gamma_{J1}}}
{{\gamma_{J0}}} - \frac{{\beta_1 }}
{{\beta_0 }}} \right) ,
\nonumber 
\eey
corresponding to 
the $\mu'$-dependent factors in \eq{evolF2}, 
forms the renormalization-group-invariant combination,
for which the Borel analysis has been performed in
the literature~\cite{neubert,bagan,penin}:
substituting \eq{fsrRG} into the RHS of \eq{rginvariant}, 
we obtain the sum rule formula for the renormalization-group-invariant 
decay constant,
\bey
&&
\!\!\!\!\!\!\!
\hat{F}^2 
e^{-\bar\Lambda/M}
\nonumber\\
&&
\!\!\!\!\!\!
=\alpha _s (\mu')^{-\frac{\gamma_{J0}}{\beta_0}} 
\left\{
\frac{N_cM^3}{\pi^2}\int_0^{\omega_{\rm th}/M}dz z^2 e^{-z}\right.
\cr&&
\!\!\!
\times
\left[1+\frac{3C_F\alpha_s(\mu')}{2\pi}\left(\ln\frac{\mu'}{2Mz}+\frac{17}{6}+\frac{2\pi^2}{9}
\right) - \frac{\alpha _s (\mu' )}{4\pi}\delta \right]
\cr&&
\!\!\!\!\!\!
-\left( {\frac{{\alpha _s (\mu' )}}
{{\alpha _s (\mu )}}} \right)^{\frac{{\gamma_{q0} }}
{{2\beta _0 }}} \left[ {1 + \frac{{\alpha _s (\mu' ) - \alpha _s (\mu )}}
{{4\pi }}\frac{{\gamma _{q0} }}
{{2\beta _0 }}\left( {\frac{{\gamma _{q1} }}
{{\gamma _{q0} }} - \frac{{\beta _1 }}
{{\beta _0 }}} \right)} \right.
\nonumber\\
&&\;\;\;\;\;\;\;\;\;\;\;\;\;\;\;\;\;\;\;\;\;\;\;\;\;\;\;\;
\left. +\frac{3C_F\alpha_s(\mu')}{2\pi} - \frac{\alpha _s (\mu' )}{4\pi}\delta 
\right]
\left\langle {\bar qq} \right\rangle (\mu )
\cr&&
+\frac{1}{16M^2} \left( {\frac{{\alpha _s (\mu' )}}
{{\alpha _s (\mu )}}} \right)^{\frac{{\gamma_{\sigma 0} }}
{{2\beta _0 }}}\mixqn (\mu )
\cr&&
\left. +\frac{\pi C_F \alpha_s(\mu)}{72N_c M^3} \qcon^2(\mu) \right\},
\label{fsrRGi}
\eey
where $\mu'=2M$ as the default scale.
\begin{figure}
\includegraphics[width=8.5cm,keepaspectratio]{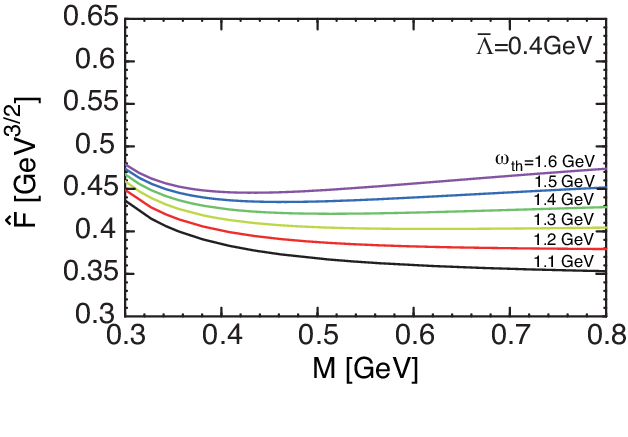}
\caption{
Borel sum rule for $\hat{F}$
based on \eq{fsrRGi} with $\mu'=2M$ and $\bar{\Lambda}=0.4$~GeV. 
From bottom to top, the curves correspond to $\omega_{\rm th}=1.1, 1.2,1.3,1.4,1.5, 1.6$~GeV.
}
\label{fRGinv1}
\end{figure}
\begin{figure}
\includegraphics[width=8.5cm,keepaspectratio]{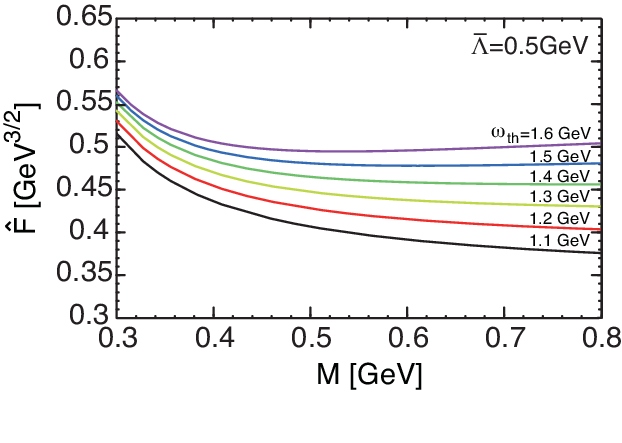}
\caption{
Borel sum rule for $\hat{F}$
based on \eq{fsrRGi} with $\mu'=2M$ and $\bar{\Lambda}=0.5$~GeV. 
From bottom to top, the curves correspond to $\omega_{\rm th}=1.1,1.2,1.3,1.4,1.5,1.6$~GeV.
}
\label{fRGinv2}
\end{figure}

In Figs.~\ref{fRGinv1} and \ref{fRGinv2}, we show $\hat{F}$ as a function of $M$ using \eq{fsrRGi}
with $\bar{\Lambda}=0.4$ and 0.5~GeV, 
respectively,
corresponding to the choice 
$\bar{\Lambda}=0.4$$-$0.5~GeV~\cite{Ball:1993xv,braun},
and the curves
are drawn for different values of $\omega_{\rm th}$.
The stable behaviors are obtained for $M \gtrsim 0.4$~GeV
with $1.2~{\rm GeV} \lesssim \omega_{\rm th}\lesssim 1.4~{\rm GeV}$,
yielding $\hat{F}\sim 0.4$~GeV$^{3/2}$.
It is worth noting that,
by taking into account the corrections due to the finite quark mass $m_b$,
the value of $\hat{F}$ around $0.4$~GeV$^{3/2}$
was shown to be modified into $f_B \simeq 0.2$~GeV~\cite{bagan,neubert0,penin}, which 
appears to be consistent with the recent lattice results, \eq{fBvalue}.
It should be noted that \eq{fsrRGi}
with $\bar{\Lambda}\gtrsim 0.5$~GeV yields the larger $\hat{F}$ as the stable value
associated with $\omega_{\rm th}\gtrsim 1.4$~GeV;
this would not be favored in view of \eq{fBvalue}.
Thus, we conclude that the range $\omega_{\rm th}=1.2$$-$1.4~GeV corresponds to 
the optimal choice for the threshold parameter~\footnote{
The sum rule analysis of the value of $\bar\Lambda$, 
calculating the ratio of \eq{fsrRGi} and its first derivative
with respect to $1/M$~\cite{bagan,neubert2},
does not serve to reduce further
the range of the values of  $\bar{\Lambda}$ nor $\omega_{\rm th}$ 
for the present case: the corresponding derivative of \eq{fsrRGi}
would receive the large
$O(\alpha_s^2)$ perturbative corrections for small $M$~\cite{neubert,neubert0}
while it 
appears to be dominated by 
higher resonances and continuum contributions for moderate as well as large $M$;
such behaviors are pronounced with the present use of $\alpha_s(1~{\rm GeV})\simeq 0.47$ 
which is larger than the value of $\alpha_s$ in the literature~\cite{bagan,neubert2}.}.

With this choice of the continuum threshold, $\omega_{\rm th}=1.2$$-$1.4~GeV, we will evaluate Borel sum rules 
for $\lambda_{E,H}^2$.
For this purpose, we need not specify the explicit value of $\bar{\Lambda}$,
because Eqs.~(\ref{RGformagnetic}) and (\ref{RGforsplitting}) divided by \eq{fsrRG}
do not depend explicitly on $\bar{\Lambda}$.
By inspection of the contributions from each term in \eq{fsrRG},
we find that, for $M\gtrsim 0.4~{\rm GeV}$,
the contribution of the perturbative correction terms is less than $\sim 40$\%
in the OPE for the sum rule 
and the contribution of the nonperturbative power correction terms is much smaller.
On the other hand, for $M \lesssim 0.6~{\rm GeV}$, the contribution of the 
higher resonances and continuum contributions is less than $\sim 40$\%
of the total contribution in the dispersion relation for the sum rule.
Thus, we take $0.4~{\rm GeV} \lesssim M \lesssim 0.6~{\rm GeV}$ as the stability window
in the following calculations of Eqs.~(\ref{RGformagnetic}) and (\ref{RGforsplitting}).
The correction terms, as well as 
the higher resonances and continuum contributions,
arising in the sum rules~(\ref{RGformagnetic}), (\ref{RGforsplitting}),
are under control for this window.

\begin{figure}
\includegraphics[width=7.4cm,keepaspectratio]{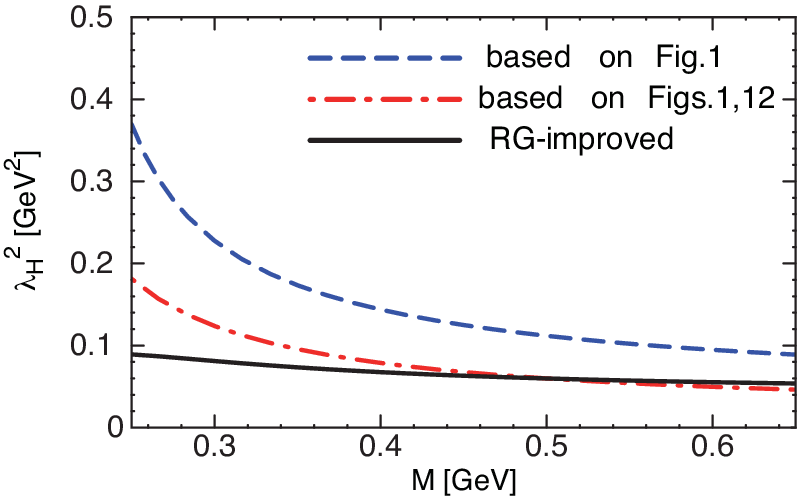}
\caption{Borel sum rules for $\lambda_{H}^2(1~{\rm GeV})$ 
with the continuum threshold $\omega_{\rm th}=1.3$~GeV.
The dashed curve is based on Fig.~\ref{treediagram},
the dot-dashed curve is based on Figs.~\ref{treediagram} and~\ref{correction-1},
and the solid curve corresponds to the renormalization-group improvement 
based on Figs.~\ref{treediagram} and~\ref{correction-1}.
}
\label{lambdah}
\end{figure}
\begin{figure}
\includegraphics[width=7.4cm,keepaspectratio]{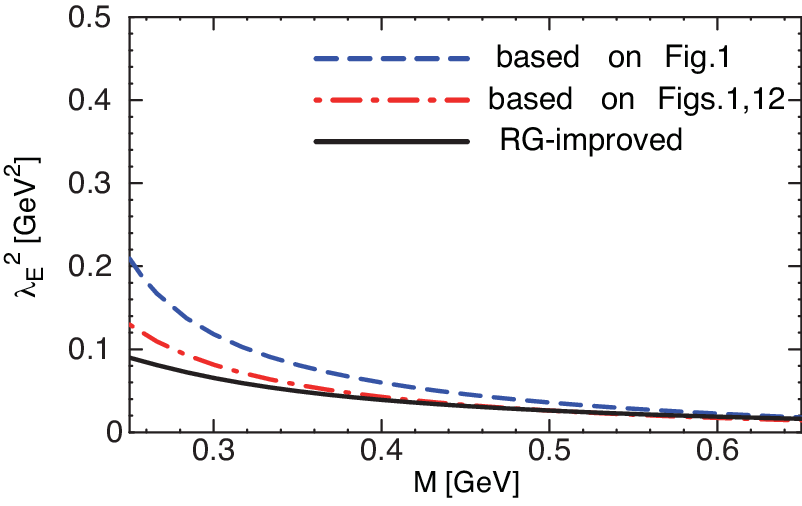}
\caption{Same as Fig.~\ref{lambdah}, but for $\lambda_{E}^2(1~{\rm GeV})$.
}
\label{lambdae}
\end{figure}
In Figs.~\ref{lambdah} and~\ref{lambdae} we show the results for $\lambda_{H}^2(\mu = 1~{\rm GeV})$
and $\lambda_{E}^2(\mu = 1~{\rm GeV})$, respectively,
as functions of the Borel parameter $M$, using the continuum threshold $\omega_{\rm th}=1.3$~GeV.
The solid curves are obtained 
by dividing Eqs.~(\ref{RGformagnetic}) and (\ref{RGforsplitting}) by \eq{fsrRG},
followed by the evolution of the results from $\mu'=2M$ to $\mu=1$~GeV using \eq{evolution},
and thus show our full results with the renormalization-group improvement. 
For comparison, the dot-dashed curves show the ``fixed-order'' results
that are obtained by dividing Eqs.~(\ref{SRformagnetic}) and (\ref{SRforsplitting}) by \eq{fsr},
while the dashed curves are obtained 
by dividing Eqs.~(\ref{treeSRformagnetic}) and (\ref{treeSRforsplitting}) by \eq{fsr0}.
The dashed curve in Fig.~\ref{lambdah} shows the behavior similar as the curves
in Fig.~\ref{mag-tree}, because the difference between those curves are only due to the value of 
the threshold $\omega_{\rm th}$.
The new $O(\alpha_s)$ contributions due to Fig.~\ref{correction-1}, 
and also the associated renormalization-group-improvement effects,
significantly improve the stability of the sum rules. 
Furthermore, those contributions significantly reduce 
the magnitude of $\lambda_H^2$ as well as $\lambda_E^2$.
For those results, we find that the behavior of the decay constant $F$,
arising in Eqs.~(\ref{RGformagnetic}) and (\ref{RGforsplitting}),
actually plays important roles: comparing Fig.~\ref{dim6-F}
with Figs.~\ref{fRGinv1},~\ref{fRGinv2}, we
see the large effects due to the renormalization-group-improved, NLO perturbative corrections 
in the decay constant sum rule.
It is worth mentioning that the corresponding large radiative corrections
to the decay constant is mainly due to one-gluon exchange between heavy and light quarks
in Feynman gauge,
i.e., from their Coulomb interaction,
and that those large effects are essential 
to yield the values consistent with \eq{fBvalue}~\cite{bagan,neubert0};
on the other hand, it has been demonstrated that
the decay constant sum rule~(\ref{fsrRG}) is quite stable with respect to inclusion of 
the NNLO-level radiative corrections~\cite{penin}.

\begin{figure}
\includegraphics[width=7.4cm,keepaspectratio]{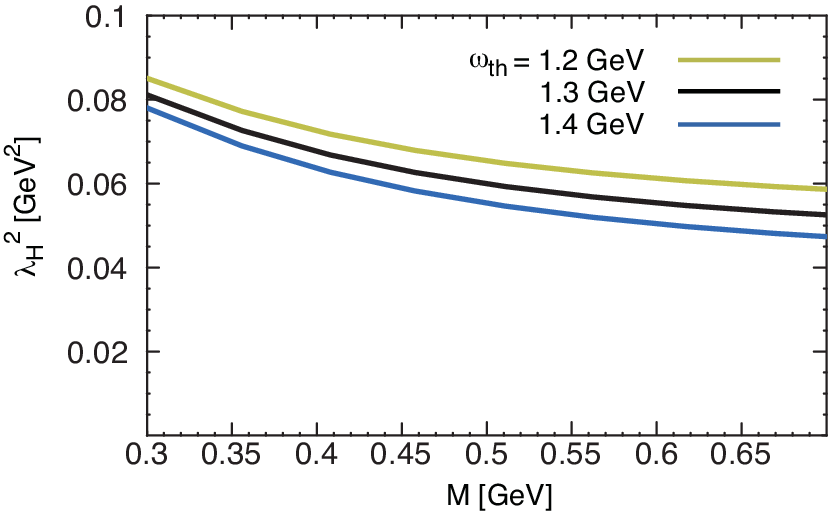}
\caption{Borel sum rules for $\lambda_{H}^2(1~{\rm GeV})$ 
as the renormalization group improvement 
based on Figs.~\ref{treediagram} and~\ref{correction-1}.
From top to bottom, the curves correspond to $\omega_{\rm th}=1.2,1.3,1.4$~GeV.
}
\label{lambdah2}
\end{figure}
\begin{figure}
\includegraphics[width=7.4cm,keepaspectratio]{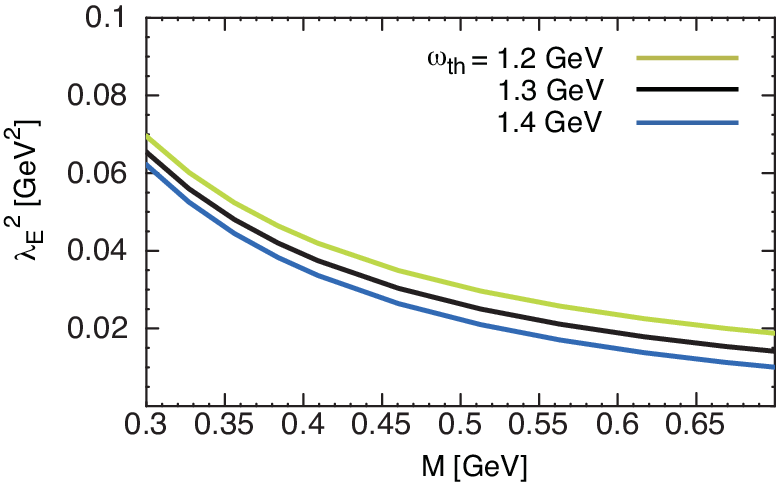}
\caption{Same as Fig.~\ref{lambdah2}, but for $\lambda_{E}^2(1~{\rm GeV})$.
}
\label{lambdae2}
\end{figure}
Figs.~\ref{lambdah2} and~\ref{lambdae2} show
the  $\omega_{\rm th}$ dependence of 
the Borel sum rules for $\lambda_{H}^2(1~{\rm GeV})$
and $\lambda_{E}^2(1~{\rm GeV})$, respectively,
as functions of $M$, 
which are obtained 
by dividing Eqs.~(\ref{RGformagnetic}) and (\ref{RGforsplitting}) by \eq{fsrRG},
followed by the evolution of the results from $\mu'=2M$ to $\mu=1$~GeV using \eq{evolution};
the middle curves in Figs.~\ref{lambdah2} and~\ref{lambdae2} are
same as the solid curves in Figs.~\ref{lambdah} and~\ref{lambdae}, respectively.
We see that the central values are $\lambda_{H}^2(1~{\rm GeV})=0.06$~GeV$^2$
and $\lambda_{E}^2(1~{\rm GeV})=0.03$~GeV$^2$
as an average over the stability window $M=0.4$$-$0.6~GeV
and the range $\omega_{\rm th}=1.2$$-$1.4~GeV for our optimized threshold parameter,
and that these central values are associated with the uncertainties
$\pm 0.01$~GeV$^2$ and $\pm 0.015$~GeV$^2$, respectively.

To estimate the uncertainties due to the lack of information of two-loop anomalous 
dimensions as well as the higher-loop effects in Eqs.~(\ref{RGformagnetic})-(\ref{evolution}),
the scale dependence of our full results with the renormalization-group improvement
is analyzed by varying $\mu'$ around the default value $2M$,
e.g., for a range $M \lesssim \mu' \lesssim 4M$; 
if $\mu'=M$ were chosen, $\alpha_s(M)$ would arise in the corresponding formulas
with too small scale for perturbation theory in 
$0.4~{\rm GeV} \lesssim M \lesssim 0.6~{\rm GeV}$.
Thus, we vary $\mu'$ in the range $1.5M \le \mu' \le 4M$,
for Eqs.~(\ref{RGformagnetic}) and (\ref{RGforsplitting}) divided by \eq{fsrRG},
followed by the evolution using \eq{evolution} from $\mu'$ to $\mu=1$~GeV:
the corresponding results of $\lambda_{E,H}^2(1~{\rm GeV})$ 
increase (decrease) for increasing (decreasing) $\mu'$, 
and
we find that 
the above-mentioned central values, $\lambda_{H}^2(1~{\rm GeV})=0.06$~GeV$^2$
and $\lambda_{E}^2(1~{\rm GeV})=0.03$~GeV$^2$, receive the 
$\pm 0.02$~GeV$^2$ and $\pm 0.005$~GeV$^2$ variations, respectively.

Similarly, we calculate the uncertainties of the results 
due to the uncertainties in the input parameters of Table~\ref{parameter}.
We also vary $\Lambda_{\rm QCD}^{(4)}$ in a range
$0.29~{\rm GeV} \lesssim \Lambda_{\rm QCD}^{(4)} \lesssim 0.33~{\rm GeV}$
corresponding to $\alpha_s(1~{\rm GeV})=0.44$$-$0.5.
Among them, the uncertainty of the dimension-5 quark-gluon-mixed condensate $\mixqn$
produces the largest effect as $\sim 15$\% and $\sim 30$\% of the total 
contribution to $\lambda_{H}^2(1~{\rm GeV})$
and $\lambda_{E}^2(1~{\rm GeV})$, respectively,
while each of the other uncertainties yields 10\% or less
of the total contribution to $\lambda_{E,H}^2(1~{\rm GeV})$.

Adding the errors induced by all source of uncertainties, $\omega_{\rm th}$, $\mu'$, condensates in Table~\ref{parameter},
and $\Lambda_{\rm QCD}^{(4)}$, discussed so far in quadrature,
we obtain
$\pm 0.025$~GeV$^2$ and $\pm 0.018$~GeV$^2$ for $\lambda_{H}^2(1~{\rm GeV})$
and $\lambda_{E}^2(1~{\rm GeV})$, respectively.
There also exists an overall intrinsic uncertainty 
of the QCD sum rule method itself which is difficult to estimate.
Thus, with a conservative estimate of the uncertainties, 
our final results read
\begin{eqnarray}
&&\lambda_E^2 (1~{\rm GeV}) = 0.03 \pm 0.02~{\rm GeV}^2 ,
\nonumber\\
&&\lambda_H^2 (1~{\rm GeV}) = 0.06 \pm 0.03~{\rm GeV}^2 .
\label{lambdaEHnew}
\end{eqnarray}
Note that the errors in the previous estimate~(\ref{lambdaEH})~\cite{GN}
are due only to the choice of the continuum threshold and
the dependence on the Borel parameter in the corresponding sum rules
using Eqs.~(\ref{treeSRformagnetic}) and (\ref{treeSRforsplitting}).

It would be interesting to compare our estimate (\ref{lambdaEHnew}) with the values of 
the corresponding quantities of light pseudoscalar mesons, $\pi$, $K$.
A straightforward comparison discussed in Appendix~\ref{compari}
suggests that the values of the quark-antiquark-gluon three-body components 
have important difference between 
the $B$ meson and the light $\pi, K$ mesons,
but their orders of magnitude 
are not largely different.
This difference in the values of the three-body components 
reflects the different behaviors of the corresponding sum rules, 
where the dimension-5 quark-gluon mixed condensates play dominant role 
in the heavy-quark limit 
while those play minor role near the chiral limit.

\section{Conclusions}
\label{sec6}

We have discussed the QCD sum rule calculation of 
the
HQET parameters
$\lambda_E^2$ and $\lambda_H^2$,
which represent quark-gluon three-body
components in the $B$-meson wavefunction.
We have updated the sum rules for $\lambda_{E,H}^2$
calculating 
the new higher-order contributions to the OPE for the corresponding correlator, 
i.e., the order $\alpha_s$ radiative corrections
to the Wilson coefficients associated with the dimension-5 quark-gluon mixed condensate, 
and
the power corrections
due to the dimension-6 vacuum condensates.
Combining with the similar NLO-level calculation for the decay-constant sum rule
which is consistent with the precise result from recent lattice QCD calculations,
we have constructed the Borel sum rules for $\lambda_{E,H}^2$.
We have found that 
the new order-$\alpha_s$ radiative corrections 
significantly reduce the values of $\lambda_{E,H}^2$,
and 
also make the corresponding sum rule formulas for $\lambda_{E,H}^2$
obey the correct renormalization-group equations.
The resummation of the relevant logarithms of the $b$-quark mass based on the renormalization group 
has been performed and proves to improve the stability of the corresponding Borel sum rules.
Our final results are obtained as \eq{lambdaEHnew},
where the perturbative as well as nonperturbative corrections
are under control and the various sources of errors are taken into account.

Compared with the previous estimate, \eq{lambdaEH}, obtained in \cite{grozin},
the central values of our results~(\ref{lambdaEHnew}) are smaller by $1/3$
and the errors are also reduced considerably.
On the other hand, the upper bounds of our results~(\ref{lambdaEHnew}) 
are close to the lower bounds of \eq{lambdaEH}.
Study of the $B$-meson light-cone distribution amplitudes using the new results
in the present paper will
be presented elsewhere.

\begin{acknowledgments}
This work was supported by the Grant-in-Aid for Scientific Research 
No.~B-19340063. 
The work of K.T. was supported in part by
the Grant-in-Aid for Scientific Research on Priority Areas
No.~22011012 and the Grant-in-Aid for Scientific Research 
Nos.~23540292, 24540284 and 25610058.
\end{acknowledgments} 
\appendix
\section{One-loop renormalization of dimension-5 heavy-light operators}
\label{operatorrenorm}

In this Appendix, we calculate the one-loop corrections
for the quark-gluon current operator arising in \eq{corfun}, 
\begin{equation}
\overline{q}gG_{\mu\nu}  \Gamma_1 h_v ,
\label{qgc}
\end{equation}
in the HQET in $D=4+2\eps$ dimensions,
and determine the corresponding renormalization constants 
in the $\overline{\rm MS}$ scheme.
Based on this result, we also write down the explicit formula implied by the ellipses
in the counter-term contribution~(\ref{ct}).

\begin{figure}
\includegraphics[width=8cm,keepaspectratio]{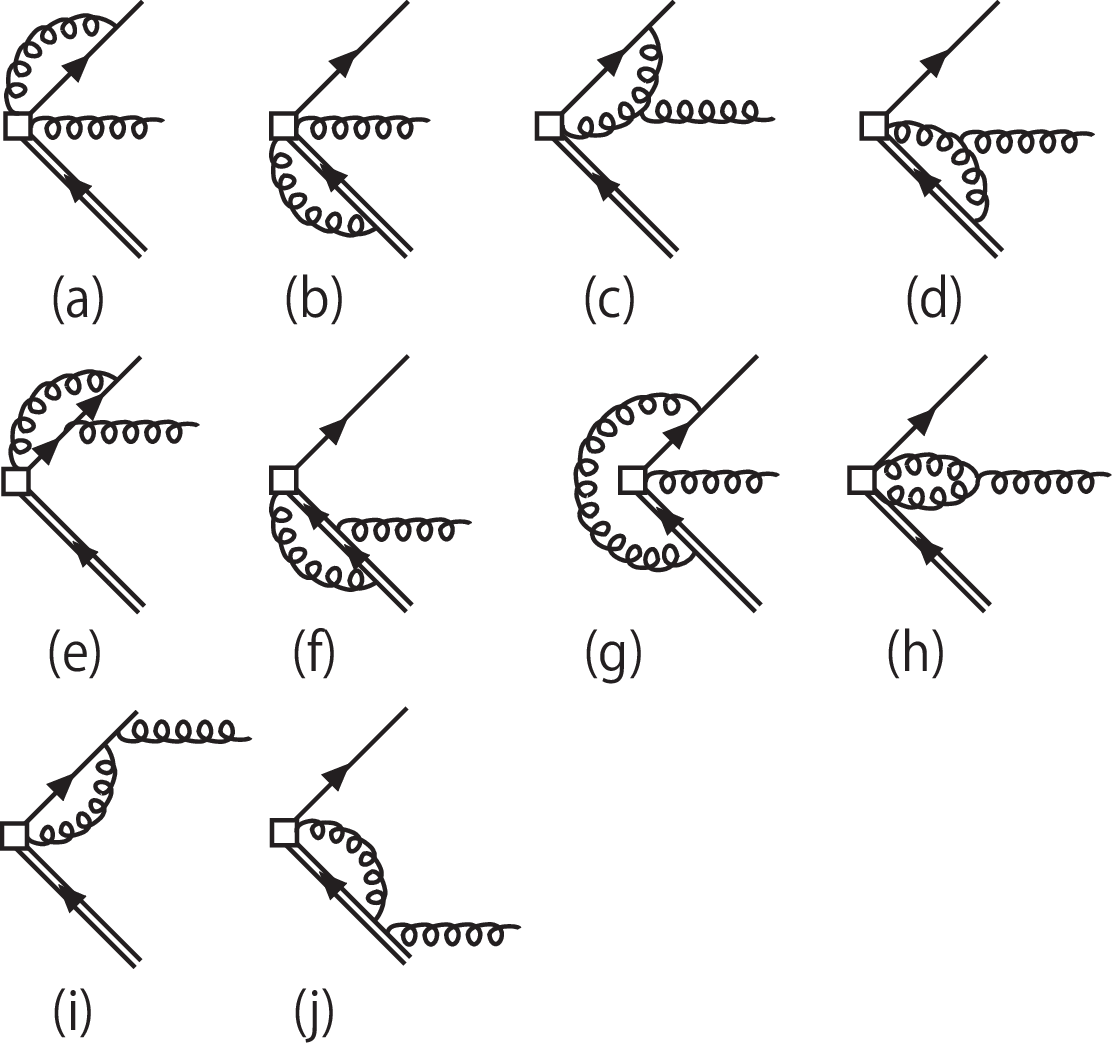}
\caption{The Feynman diagrams relevant for the one-loop renormalization
of the quark-gluon three-body operator (\ref{qgc}).}
\label{3body}
\end{figure}

In \eq{qgc}, $\Gamma_1$ is an arbitrary gamma matrix,
and $\mu$, $\nu$ are the free Lorentz indices.
The relevant one-loop diagrams
are shown in Fig.~\ref{3body},
and those loop corrections for the operator~(\ref{qgc})
induce the mixing
with the dimension-5 heavy-light operators having the same
Lorentz-transformation property as that of \eq{qgc}.
The corresponding mixing matrix 
can be obtained,
in principle, by calculating only the one-particle irreducible diagrams~(a)-(h) in Fig.~\ref{3body},
but it is known, for the case of the higher dimensional operators like \eq{qgc},
that such procedure requires to treat explicitly the additional mixing of the 
operators that would vanish by the use of the equations of motion~\cite{Politzer:1980me};
in particular, this implies that we have to use the off-shell external fields
which, in turn, allow the mixing of the gauge-noninvariant operators as well as the gauge-invariant ones,
via the corresponding one-particle irreducible diagrams,
as demonstrated in many works~\cite{kodaira} for the case of the renormalization of the higher twist operators relevant to the 
nucleon structure functions.

Here, to avoid such complication associated with the mixing of those ``alien'' operators,
we calculate the relevant one-particle-reducible diagrams~(i), (j) in Fig.~\ref{3body} as well,
so that we can adopt the usual background field method, as in the calculations 
of the correlators discussed in the main text;
then, we may use the equations of motion for the external fields at any step of calculation~\cite{Kodaira:1997ig}
and the contribution from each diagram of Fig.~\ref{3body} is obtained 
in a gauge-invariant form
expressed 
solely in terms of 
the (many) 
three-body operators 
of the similar
type 
as \eq{qgc}.
We use the building blocks (\ref{hprop}), (\ref{qprop}), (\ref{gprop}), etc., and 
carry out
the loop integrations in the coordinate space.
The contributions of the diagrams~(f) and (j) in Fig.~\ref{3body} vanish because
these diagrams contain the vanishing subdiagrams~(a) and (b) in Fig.~\ref{subdiagram}, respectively.
Also, the diagrams~(a) and (b) in Fig.~\ref{3body} vanish by the reason similar as 
the diagrams in Fig.~\ref{correction-2}.
The sum of the UV poles from all the other diagrams in Fig.~\ref{3body} reads,
\bey
&& \!\!\!\!\!\!\!
\frac{\alpha_s}{4\pi\eps}\left(
-\frac{N_c}{8}\left[
\overline{q} gG_{\mu \nu}\Gamma_1 h_v 
+3i 
\overline{q} gG_{\mu\rho}{\sigma_{\nu}}^{\rho}\Gamma_1h_v 
\right]
\right.
\cr&&
\cr&&
-\frac{N_c}{2}
\overline{q} gG_{\mu\rho} v^\rho v_\nu \Gamma_1 h_v
%
\cr&&
+\frac{i}{32 N_c}
\overline{q} gG^{\rho\lambda} \g_\mu\left[\g_\nu \sig_{\rho\lambda}+\sig_{\rho\lambda}\g_\nu
\right] \Gamma_1 h_v 
\cr&&
+\left[\frac{1}{4N_c}+N_c\right]
\overline{q} gG_{\mu\nu} \Gamma_1 h_v
\cr&&
+\frac{C_F}{24}\left[ 2\overline{q} gG_{\mu\nu}\Gamma_1 h_v + \overline{q} gG\cdot \sig \sig_{\mu\nu}\Gamma_1 h_v
\right.
\cr&& \;\;\;\;\;\;\;\;
\left.\left.
+2i
\overline{q} gG_{\mu\rho}{\sig_\nu}^\rho \Gamma_1 h_v
\right]\right)
\cr&&
\!-\left( \mu\leftrightarrow\nu \right) ,
\label{apparesult}
\eey
with the last line indicating that all the preceding terms have to be antisymmetrized
under the interchange $\mu\leftrightarrow\nu$.
Here, the first three lines correspond to the contributions
of the diagrams~(c), (d), and (e) in Fig.~\ref{3body},
the fourth line corresponds to the diagrams~(g) and (h),
and the fifth and sixth lines correspond to the diagram (i).

Now, the counter-term contribution for the renormalization of the operator~(\ref{qgc})
should be combined with \eq{apparesult},
so as to cancel all the UV poles arising in \eq{apparesult}.
Thus, the corresponding counter-term contribution
is given by the minus of \eq{apparesult} with the replacement, 
$1/\eps \rightarrow 1/\hat{\eps}$ ($=1/\eps +\gamma_E -\ln4\pi$),
in the $\overline{\rm MS}$ scheme.
Substituting this counter-term contribution for the quark-gluon 
current~(\ref{qgc}) into \eq{corfun} and evaluating the corresponding correlator in $D$ dimensions under the background 
fields, 
we immediately find that the formula implied by the ellipses in \eq{ct}
is formally given by \eq{Phi} with $\overline{q}(0) gG_{\mu\nu}(0)$
replaced by the minus of \eq{apparesult} with $h_v \rightarrow 1$, $\eps \rightarrow \hat{\eps}$, 
and reads, using $\widehat{\Pi}^{\rm tree}_{3H}(\omega)$ of \eq{pitree},
\bey
&& \!\!\!\!\!\!
\frac{\alpha_s}{8\pi\hat{\eps}} 
\Tr\left[\left\{
-\frac{N_c}{8}\left( \sig_{\mu \nu} +3 i\sig_{\mu\rho}{\sigma_{\nu}}^{\rho} \right)
\right.\right.
\nonumber\\
&&
-\frac{N_c}{2}
\sig_{\mu\rho} v^\rho v_\nu 
\nonumber\\
&&
+\frac{1}{32 N_c}
i\sig^{\rho\lambda} \g_\mu\left(\g_\nu \sig_{\rho\lambda}+\sig_{\rho\lambda}\g_\nu
\right)
\nonumber\\
&&
+\left(\frac{1}{4N_c} + N_c \right) \sig_{\mu \nu} 
\nonumber\\&&
\!\!\!\!\!\!
\left.\left.  
+ \frac{C_F}{24}\!
\left(2\sig_{\mu \nu} \!+\! \left(\sig_{\rho \lambda}\right)^2 \sig_{\mu\nu}
\!+\!2i\sig_{\mu\rho}{\sigma_{\nu}}^{\rho}\right)
\right\}\! \Gamma_1  \pp \Gamma_2 \right]\! \widehat{\Pi}^{\rm tree}_{3H}(\omega)
\nonumber\\&&
\!\!\!\!-\left( \mu\leftrightarrow\nu \right) ,
\nonumber\\
&&
\!\!\!\!\!
=\frac{\alpha_s}{8\pi} 
\left\{\! \left( \frac{1}{\hat{\eps}}\left[ N_c-\frac{1}{2N_c} \right]
-\frac{3N_c}{4}-\frac{1}{2N_c} \right) \! \Tr\left[\sig_{\mu\nu}\Gamma_1 \pp \Gamma_2 \right]
\right.
\nonumber\\
&&
\left.+\frac{N_c}{2\hat{\eps}} \Tr\left[(iv_\mu \gamma_\nu -iv_\nu \gamma_\mu)\Gamma_1 \pp \Gamma_2 \right]
\right\} \widehat{\Pi}^{\rm tree}_{3H}(\omega) ,
\label{ctct}
\eey
where the final form in the RHS is presented up to the irrelevant terms that vanish as $\eps \rightarrow 0$.

It is worth mentioning that
the results based on \eq{apparesult} with the particular choices, 
$\Gamma_1=\sigma^{\mu \nu} \gamma_5$ and $\Gamma_1= \g^\mu\gamma_5 v^\nu$,
reproduce the one-loop renormalization mixing matrix between the two pseudoscalar operators,
$\overline{q} gG\cdot \sig \gamma_5h_v$ and $\overline{q} gG_{\mu \nu} \g^\mu v^\nu \gamma_5 h_v$, calculated in \cite{Grozin:1996hk}.
In particular, it is straightforward to see that the matrix elements of a 
system of the corresponding operator-mixing formulas
yield the one-loop renormalization group equations 
for $\lambda^2_{E,H}$ of Eqs.~(\ref{lame}),~(\ref{lamh}), as~\cite{Grozin:1996hk}
\begin{equation}
\mu \frac{d}{d\mu} 
\left( 
\begin{array}{c}
\lambda_E^2(\mu)\\
\lambda_H^2(\mu)
\end{array}
\right) 
+ \frac{\alpha_s(\mu)}{4\pi} \hat{\gamma}_0 
\left( 
\begin{array}{c}
\lambda_E^2(\mu)\\
\lambda_H^2(\mu)
\end{array}
\right) =0 ,
\label{elam}
\end{equation}
with the mixing matrix,
\begin{equation}
\hat{\gamma}_0 = 
\left( 
\begin{array}{cc}
\frac{8}{3} C_F+\frac{3}{2}N_c &\;\;\;\;\; \frac{4}{3}C_F  -\frac{3}{2}N_c\\
\frac{4}{3}C_F  -\frac{3}{2}N_c&\;\;\;\;\; \frac{8}{3}C_F  +\frac{5}{2}N_c
\end{array}
\right) .
\label{mm}
\end{equation}

Finally, we mention that the one-loop renormalization of the operator~(\ref{calo})
to obtain the renormalization constant~(\ref{zo}) can be carried out in a similar manner as above.
Indeed, most results can be obtained from the above result (\ref{apparesult}) by the formal substitutions,
$h_v \rightarrow q$ and $\Gamma_1 \rightarrow \sig_{\mu\nu}$, for an external quark field 
and the gamma matrix structure, respectively,
and thus need not new calculation.
The only diagram that requires new calculation is the one corresponding to the diagram~(g) in Fig.~\ref{3body}.
We have only one flavor-singlet, scalar three-body operator of dimension-5, given by \eq{calo},
and thus do not encounter the operator mixing in the background field method for this case.

\section{Quark-gluon three-body components in the $\pi$ and $K$ mesons}
\label{compari}

In this Appendix, we compare our estimate (\ref{lambdaEHnew}) 
of the quark-gluon three-body components in the $B$ meson
with the values of the corresponding quantities of light pseudoscalar mesons, $\pi$, $K$.
For definiteness,
we will give most of the following discussions for the case of $K$ mesons, i.e., $s\bar{q}$ bound states with $q=u,d$.
The $K$-meson matrix elements of three-body local operators of dimension 5 in QCD
read
\bey
\bra 0|\overline{q}\gamma_{\alpha}g\widetilde{G}_{\rho\sigma} s|K(p)\ket&=&\frac{i}{3}f_K\delta_K^2( g_{\alpha\sigma}p_\rho-
g_{\alpha\rho}p_\sigma),\label{lightmesonmat00}\\
\bra 0|\overline{q}\gamma_{\alpha}\gamma_5 ig G_{\rho\sigma} s|K(p)\ket&=&\frac{i}{3}f_K m_K^2
\kappa_{4K} (g_{\alpha\sigma}p_\rho-g_{\alpha\rho}p_\sigma),\label{lightmesonmat0}\\
\bra 0|\overline{q}\sigma_{\alpha\beta}\gamma_5 gG_{\mu\nu} s|K(p)\ket&=&if_{3K}\left[p_\beta\left(g_{\alpha\mu}p_\nu-g_{\alpha\nu}p_\mu\right)-p_\alpha\left(g_{\beta\mu}p_\nu-g_{\beta\nu}p_\mu\right)
\right]+i m_K^2 \varphi_K \left(g_{\alpha\mu}g_{\beta \nu}
-g_{\alpha\nu}g_{\beta \mu} \right) ,
\label{lightmesonmat}
\eey
where $\widetilde{G}_{\rho\sigma}=\frac{1}{2}\epsilon_{\rho\sigma\xi\eta}G^{\xi\eta}$, 
$|K(p)\ket$ is the kaon state
with the 4-momentum $p$, obeying $\bra K(p) |K(p')\ket = 2 p^0 (2\pi)^3 \delta^{(3)}(p-p')$ and $p^2=m_K^2$,
and $f_K$ denotes the decay constant, as usual,
\begin{equation}
\bra 0|\overline{q}\gamma_\rho \g_5s |K(p)\ket
=if_K p_\rho .
\label{fK}
\end{equation}
We follow the notation of Ref.\cite{BBL}
for the nonperturbative parameters 
$f_{3K}$,  $\delta_K^2$,  and $\kappa_{4K}$; 
$f_{3K}$ corresponds to twist-3 and $\delta_K^2$,  $\kappa_{4K}$ correspond to  twist-4.
Here, we also need to deal explicitly with $\varphi_K$, which corresponds to twist-5.
The matrix elements of $\overline{q}Gs$ operators associated with the Dirac matrices other than those 
in Eqs.~(\ref{lightmesonmat00})-(\ref{lightmesonmat}) vanish.
Thus, the meson matrix elements of the dimension-5 three-body operators in QCD
are expressed by the four independent nonperturbative parameters; compare with the corresponding
$B$-meson matrix element (\ref{mat3}) expressed by the two independent parameters $\lambda_{E,H}^2$.
In the rest frame, Eqs.~(\ref{lightmesonmat00})-(\ref{lightmesonmat}) yield
the matrix elements associated with the chromoelectric and chromomagnetic fields, as 
\begin{eqnarray}
&&\bra 0|\overline{q}\bfm{\alpha}\cdot g\bfm{E}\g_5\gamma_0s|K(p)\ket=-m_K^3 f_K \kappa_{4K} ,
\nonumber \\
&&\bra 0|\overline{q}\bfm{\sigma}\cdot g\bfm{H}\g_5\gamma_0s| K(p)\ket=i m_K f_K \delta_K^2 ,
\nonumber \\
&&\bra 0|\overline{q}\bfm{\alpha}\cdot g\bfm{E}\g_5s| K(p)\ket=-3m_K^2 \left( f_{3K}+\varphi_K \right) ,
\nonumber
\\
&&\bra 0|\overline{q}\bfm{\sigma}\cdot g\bfm{H}\g_5s| K(p)\ket=-3i m_K^2 \varphi_K,
\label{Kh} 
\end{eqnarray}
with 
$p=(m_K, \bfm{0})$, and, using \eq{fK}, 
the ratios independent of the normalization of the meson state read,
\begin{eqnarray}
&&\frac{i\bra 0|\overline{q}\bfm{\alpha}\cdot g\bfm{E}\g_5\gamma_0s|K(p)\ket}{\bra 0|\overline{q}\gamma_0 \g_5s |K(p)\ket}=-m_K^2 \kappa_{4K} ,
\label{KKe0}\\
&&\frac{\bra 0|\overline{q}\bfm{\sigma}\cdot g\bfm{H}\g_5\gamma_0s| K(p)\ket}{\bra 0|\overline{q}\gamma_0 \g_5s |K(p)\ket}
= \delta_K^2 ,
\label{KKh0}\\
&&\frac{\bra 0|\overline{q}\bfm{\sigma}\cdot g\bfm{H}\g_5s| K(p)\ket-i\bra 0|\overline{q}\bfm{\alpha}\cdot g\bfm{E}\g_5s| K(p)\ket}
{\bra 0|\overline{q}\gamma_0 \g_5s |K(p)\ket}=3m_K \frac{f_{3K}}{f_K},
\label{KKeh}
\end{eqnarray}
where the last formula is obtained 
by eliminating the twist-5 quantity $\varphi_K$, whose value is unknown.
On the other hand, using Eqs.~(\ref{fmu}), (\ref{lame}) and  (\ref{lamh}),
we have the similar ratios for the $B$-meson at rest:
\begin{eqnarray}
&&\frac{i\bra 0|\overline{q}\bfm{\alpha}\cdot g\bfm{E}\g_5h_v|\bar B(v)\ket}{\bra 0|\overline{q}\gamma_0 \g_5h_v |\bar B(v)\ket}
=\lambda_E^2,
\label{lameratio}
\\
&&\frac{\bra 0|\overline{q}\bfm{\sigma}\cdot g\bfm{H}\g_5h_v|\bar B(v)\ket}{\bra 0|\overline{q}\gamma_0 \g_5h_v |\bar B(v)\ket}
= \lambda_H^2.
\label{lamhratio} 
\end{eqnarray}
Therefore, 
the value of $\lambda_H^2 - \lambda_E^2$ may be compared with that of \eq{KKeh}.
The use of $h_v=\Slash{v}h_v=\gamma_0h_v$ in the rest frame
would suggest another (rough) comparison of \eqs{lameratio}{lamhratio}, respectively, with \eqs{KKe0}{KKh0}.

In Ref.\cite{BBL}, numerical values of the nonperturbative parameters $\delta_K^2, \kappa_{4K}$ and $f_{3K}$ 
and the similar quantities for the pion are evaluated from QCD sum rules,
using the QCD correlation functions analogous to the HQET correlators treated in this paper~\footnote{Estimate of the 
nonperturbative parameters in Ref.\cite{BBL} is performed using
not only the correlators between the two- and three-body currents but also those between
the two three-body currents.}.
The values presented in Ref.\cite{BBL}
are
\bey
&&\delta_\pi^2=0.18\pm0.06~{\rm GeV}^2,\quad \kappa_{4\pi}=0, \quad\quad\quad\quad \quad\quad
f_{3\pi}=0.0045\pm0.0015~{\rm GeV}^2,\cr
&&\delta_K^2=0.20\pm0.06~{\rm GeV}^2,\quad  \kappa_{4K}=-0.09\pm0.02,  \quad f_{3K}=0.0045\pm0.0015~{\rm GeV}^2,
\label{piKconst}
\eey
for the $\pi$ as well as $K$ meson, where all values are given at the scale $\mu=1$~GeV.
Note that the G-parity-breaking contribution (\ref{lightmesonmat0}) vanishes for the pion case. 
Combined with $m_\pi=140~{\rm MeV}$, $m_K=494~{\rm MeV}$, $f_\pi=131~{\rm MeV}$, 
and $f_K=160~{\rm MeV}$, we obtain the values of Eqs.~(\ref{KKe0})-(\ref{KKeh}):
\begin{eqnarray}
&&\frac{\bra 0|\overline{q}\bfm{\sigma}\cdot g\bfm{H}\g_5s| K(p)\ket-i\bra 0|\overline{q}\bfm{\alpha}\cdot g\bfm{E}\g_5s| K(p)\ket}
{\bra 0|\overline{q}\gamma_0 \g_5s |K(p)\ket}=0.042\pm 0.014~{\rm GeV}^2,\nonumber\\ 
&&\frac{i\bra 0|\overline{q}\bfm{\alpha}\cdot g\bfm{E}\g_5\gamma_0s|K(p)\ket}{\bra 0|\overline{q}\gamma_0 \g_5s |K(p)\ket}=0.022\pm 0.005~{\rm GeV}^2,\quad\quad
\frac{\bra 0|\overline{q}\bfm{\sigma}\cdot g\bfm{H}\g_5\gamma_0s| K(p)\ket}{\bra 0|\overline{q}\gamma_0 \g_5s |K(p)\ket}=0.20\pm 0.06~{\rm GeV}^2,
\label{esK}
\end{eqnarray}
which may be compared with our values of  $\lambda_H^2 - \lambda_E^2$, $\lambda_E^2$, and $\lambda_H^2$, respectively,
see \eq{lambdaEHnew}.
For the pion, we obtain similarly, 
\begin{eqnarray}
&&\frac{\bra 0|\overline{q}\bfm{\sigma}\cdot g\bfm{H}\g_5s| \pi(p)\ket-i\bra 0|\overline{q}\bfm{\alpha}\cdot g\bfm{E}\g_5s| \pi(p)\ket}
{\bra 0|\overline{q}\gamma_0 \g_5s |K(p)\ket}=0.014 \pm 0.005~{\rm GeV}^2,\nonumber\\
&&\frac{i\bra 0|\overline{q}\bfm{\alpha}\cdot g\bfm{E}\g_5\gamma_0s|\pi(p)\ket}{\bra 0|\overline{q}\gamma_0 \g_5s |\pi(p)\ket}=0,\quad\quad\quad
\frac{\bra 0|\overline{q}\bfm{\sigma}\cdot g\bfm{H}\g_5\gamma_0s| \pi(p)\ket}{\bra 0|\overline{q}\gamma_0 \g_5s |\pi(p)\ket}=0.18\pm 0.06~{\rm GeV}^2.
\label{espi}
\end{eqnarray}
Thus, the quark-antiquark-gluon three-body components 
are different between 
the $B$ meson and the light $\pi, K$ mesons, but 
the present results suggest that their orders of magnitude are not largely different.
In particular, the ``splitting'' between the chromomagnetic and chromoelectric fields, $\lambda_H^2 - \lambda_E^2$, 
has similar size as the first quantity in 
\eqs{esK}{espi}. In this connection,
it is worth mentioning the following:
the dimension-5 quark-gluon mixed condensates arise accompanying a quark mass, as
$m_q\mixqn$, $m_s \mixqn$, etc., in the OPE in full QCD to derive the sum rules in Ref.\cite{BBL}, so that
the mixed condensates play minor role near the chiral limit.
On the other hand, the quark-gluon mixed condensates give a dominant contribution 
in the heavy-quark limit for the $B$-meson case, as demonstrated in this paper, but the splitting $\lambda_H^2 - \lambda_E^2$
receives the contribution of the mixed condensates at $O(\alpha_s)$, 
see \eq{SRforsplitting}.

\def\Ref#1{[\ref{#1}]}
\def\Refs#1#2{[\ref{#1},\ref{#2}]}
\def\npb#1#2#3{{Nucl. Phys.\,}{\bf B{#1}},\,#2\,(#3)}
\def\npa#1#2#3{{Nucl. Phys.\,}{\bf A{#1}},\,#2\,(#3)}
\def\np#1#2#3{{Nucl. Phys.\,}{\bf{#1}},\,#2\,(#3)}
\def\plb#1#2#3{{Phys. Lett.\,}{\bf B{#1}},\,#2\,(#3)}
\def\prl#1#2#3{{Phys. Rev. Lett.\,}{\bf{#1}},\,#2\,(#3)}
\def\prd#1#2#3{{Phys. Rev.\,}{\bf D{#1}},\,#2\,(#3)}
\def\prc#1#2#3{{Phys. Rev.\,}{\bf C{#1}},\,#2\,(#3)}
\def\prb#1#2#3{{Phys. Rev.\,}{\bf B{#1}},\,#2\,(#3)}
\def\pr#1#2#3{{Phys. Rev.\,}{\bf{#1}},\,#2\,(#3)}
\def\ap#1#2#3{{Ann. Phys.\,}{\bf{#1}},\,#2\,(#3)}
\def\prep#1#2#3{{Phys. Reports\,}{\bf{#1}},\,#2\,(#3)}
\def\rmp#1#2#3{{Rev. Mod. Phys.\,}{\bf{#1}},\,#2\,(#3)}
\def\cmp#1#2#3{{Comm. Math. Phys.\,}{\bf{#1}},\,#2\,(#3)}
\def\ptp#1#2#3{{Prog. Theor. Phys.\,}{\bf{#1}},\,#2\,(#3)}
\def\ib#1#2#3{{\it ibid.\,}{\bf{#1}},\,#2\,(#3)}
\def\zsc#1#2#3{{Z. Phys. \,}{\bf C{#1}},\,#2\,(#3)}
\def\zsa#1#2#3{{Z. Phys. \,}{\bf A{#1}},\,#2\,(#3)}
\def\intj#1#2#3{{Int. J. Mod. Phys.\,}{\bf A{#1}},\,#2\,(#3)}
\def\sjnp#1#2#3{{Sov. J. Nucl. Phys.\,}{\bf #1},\,#2\,(#3)}
\def\pan#1#2#3{{Phys. Atom. Nucl.\,}{\bf #1},\,#2\,(#3)}
\def\app#1#2#3{{Acta. Phys. Pol.\,}{\bf #1},\,#2\,(#3)}
\def\jmp#1#2#3{{J. Math. Phys.\,}{\bf {#1}},\,#2\,(#3)}
\def\cp#1#2#3{{Coll. Phen.\,}{\bf {#1}},\,#2\,(#3)}
\def\epjc#1#2#3{{Eur. Phys. J.\,}{\bf C{#1}},\,#2\,(#3)}
\def\mpla#1#2#3{{Mod. Phys. Lett.\,}{\bf A{#1}},\,#2\,(#3)}
\def\etal{{\it et al.}}

\end{document}

%% file: def.tex
\def\ben{\begin{equation}}
\def\een{\end{equation}}
\def\bey{\begin{eqnarray}}
\def\eey{\end{eqnarray}}
\def\ba{\begin{array}}
\def\ea{\end{array}}
\def\benmrt{\begin{enumerate}}
\def\eenmrt{\end{enumerate}}

\def\psla{p{\raise1pt\hbox{$\!\!\!/$}}}
\def\dsla{\partial{\raise1pt\hbox{$\!\!\!/$}}}
\def\Dsla{D{\raise1pt\hbox{$\!\!\!\!/$}}}
\def\xsla{x{\raise1pt\hbox{$\!\!\!/$}}}

\def\jmu5{j_{\mu 5}^{(i)}(0)}
\def\jnu5{j_{\nu 5}^{(i)}(0)}

\def\qq0v{\langle0\!\mid\!{\bar q}q\!\mid\! 0\rangle}

\def\qc0f{\langle0\!\mid\!{\bar q}q\!\mid\!0\rangle_{F}}

\def\qsq0f{\langle0\!\mid\!{\bar q}\sigma_{\mu\nu}q\!\mid\!0\rangle_{F}}

\def\qgdq0f{\langle0\!\mid\!{\bar q}{\cal 
S}\gamma_{\mu}D_{\nu}q\!\mid\!0\rangle_{F}}

\def\qddq0f{\langle0\!\mid\!{\bar q}{\cal
S}D_{\mu}D_{\nu}q\!\mid\!0\rangle_{F}}

\def\3mmtm{|{\bf q}|^2}

\def\eq#1{Eq.(\ref{#1})}
\def\eqs#1#2{Eqs.(\ref{#1}) and (\ref{#2})}

\def\Ref#1{[\ref{#1}]}
\def\Refs#1#2{[\ref{#1},\ref{#2}]}

\def\p0{p_0}

\def\gam3{\mbox{\boldmath{$\gamma$}}}
\def\e0{E_{0}(s_{0},s)}
\def\e1{E_{1}(s_{0},s)}
\def\e2{E_{2}(s_{0},s)}

\def\aplt{\kern0.3333em \raise 0.2ex \hbox{$<$}%
\kern-0.8em \lower0.8ex \hbox{$\sim$}%
\kern0.3333em}
\def\aplg{\kern0.3333em \raise 0.2ex \hbox{$>$}%
\kern-0.8em \lower0.8ex \hbox{$\sim$}%
\kern0.3333em}

\def\ln{{\rm ln}}

%% file: finalv3.bbl
\begin{thebibliography}{99}
\bibitem{Antonelli:2009ws}
  M.~Antonelli {\it et al.},
  Phys.\ Rept.\  {\bf 494}, 197 (2010).
\bibitem{neubert0}
M.~Neubert,
Phys.\ Rept.\ {\bf 245}, 259 (1994).
\bibitem{Gamiz:2009ku}
  E.~Gamiz, C.~T.~H.~Davies, G.~P.~Lepage, J.~Shigemitsu and M.~Wingate
                  [HPQCD Collaboration],
  Phys.\ Rev.\ {\bf D80}, 014503 (2009).
\bibitem{Bernard}
 C.~Bernard {\it et al.},
  PoS {\bf LATTICE2008}, 278 (2008);
A.~Bazavov {\it et al.}  [Fermilab Lattice and MILC Collaborations],
  Phys.\ Rev.\ {\bf D85}, 114506 (2012).
\bibitem{Blossier:2009hg} 
  B.~Blossier {\it et al.}  [ETM Collaboration],
  JHEP {\bf 1004}, 049 (2010).
\bibitem{Schwartz:2009hv}
  A.~J.~Schwartz,
  AIP Conf.\ Proc.\  {\bf 1182}, 299 (2009).
\bibitem{Hara:2010dk}
K.~Ikado {\it et al.}  [Belle Collaboration],
  Phys.\ Rev.\ Lett.\  {\bf 97}, 251802 (2006);
 K.~Hara {\it et al.}  [Belle Collaboration],
  Phys.\ Rev.\ {\bf D82}, 071101 (2010);
I.~Adachi {\it et al.}  [Belle Collaboration],
  Phys.\ Rev.\ Lett.\  {\bf 110}, 131801 (2013).
\bibitem{Aubert:2007bx}
  B.~Aubert {\it et al.}  [BABAR Collaboration],
  Phys.\ Rev.\ {\bf  D76}, 052002 (2007);
  P.~del Amo Sanchez {\it et al.}  [BaBar Collaboration],
  arXiv:1008.0104 [hep-ex];
B.~Aubert {\it et al.}  [BaBar Collaboration],
  Phys.\ Rev.\ D {\bf 81}, 051101 (2010).
\bibitem{GN}
A.~G.~Grozin and M.~Neubert, 
Phys.\ Rev.\ {\bf D55}, 272 (1997).
\bibitem{bagan}
  E.~Bagan, P.~Ball, V.~M.~Braun and H.~G.~Dosch,
  Phys.\ Lett.\ {\bf B278}, 457 (1992).
\bibitem{penin}
  A.~A.~Penin and M.~Steinhauser,
  Phys.\ Rev.\ {\bf D65}, 054006 (2002).
\bibitem{zzc}
  A.~R.~Zhitnitsky, I.~R.~Zhitnitsky and V.~L.~Chernyak,
  Sov.\ J.\ Nucl.\ Phys.\  {\bf 41}, 284 (1985)
  [Yad.\ Fiz.\ {\bf 41}, 445 (1985)];
  V.~M.~Braun and I.~E.~Filyanov,
  Z.\ Phys.\ {\bf  C48}, 239 (1990);
P.~Ball, V.~M.~Braun, Y.~Koike and K.~Tanaka,
  Nucl.\ Phys.\ {\bf B529}, 323 (1998).
\bibitem{KKQT}
  H.~Kawamura, J.~Kodaira, C.F.~Qiao and K.~Tanaka,
  Phys. Lett. {\bf B523}, 111 (2001);
  Erratum-ibid. {\bf B536}, 344 (2002); 
  Mod. Phys. Lett. {\bf A18}, 799 (2003); 
  Nucl. Phys. B (Proc. Suppl.) {\bf 116}, 269 (2003);
H.~Kawamura, J.~Kodaira and K.~Tanaka,
  Prog.\ Theor.\ Phys.\  {\bf 113}, 183 (2005).
\bibitem{GW}
  T.~Huang, X.~G.~Wu and M.~Z.~Zhou,
  Phys.\ Lett.\ {\bf B611}, 260 (2005); 
  B.~Geyer and O.~Witzel,
  Phys.\ Rev.\ {\bf D72}, 034023 (2005);
  T.~Huang, C.~F.~Qiao and X.~G.~Wu,
  Phys.\ Rev.\ {\bf D73}, 074004 (2006);
B.~Geyer and O.~Witzel,
  Phys.\ Rev.\ {\bf  D76}, 074022 (2007);
  A.~Khodjamirian, T.~Mannel and N.~Offen,
  Phys.\ Lett.\ {\bf B620}, 52 (2005);
  A.~Khodjamirian, T.~Mannel and N.~Offen,
  Phys.\ Rev.\ {\bf  D75}, 054013 (2007);
A.~Le Yaouanc, L.~Oliver and J.~C.~Raynal,
  Phys.\ Rev.\ {\bf D77}, 034005 (2008).
\bibitem{braun}
V.~M.~Braun, D.~Y.~Ivanov and G.~P.~Korchemsky,
  Phys. Rev. {\bf D69}, 034014 (2004).
\bibitem{Lee:2005gza}
  S.~J.~Lee and M.~Neubert,
  Phys. Rev. {\bf D72}, 094028 (2005).
\bibitem{DescotesGenon:2009hk}
  S.~Descotes-Genon and N.~Offen,
  JHEP {\bf 0905}, 091 (2009);
M.~Knodlseder and N.~Offen,
  JHEP {\bf 1110}, 069 (2011).
\bibitem{Kawamura:2008vq}
  H.~Kawamura and K.~Tanaka,
  Phys.\ Lett.\ {\bf B673}, 201 (2009);
H.~Kawamura and K.~Tanaka,
  Phys.\ Rev.\ {\bf  D81}, 114009 (2010).
\bibitem{Bell08}
  G.~Bell and V.~Pilipp,
  Phys.\ Rev.\ {\bf D80}, 054024 (2009);
  M.~Beneke, T.~Huber and X.~Q.~Li,
  Nucl.\ Phys.\ {\bf B832}, 109 (2010).
\bibitem{shuryak}
E.~V.~Shuryak,
  Nucl.\ Phys.\ {\bf B198}, 83 (1982).
\bibitem{neubert}
M.~Neubert, \prd{45}{2451}{1992}.
\bibitem{novikov}
  V.~A.~Novikov, M.~A.~Shifman, A.~I.~Vainshtein and V.~I.~Zakharov,
  Fortsch.\ Phys.\  {\bf 32}, 585 (1984).
\bibitem{grozin} See, e.g., A.~G.~Grozin,
  Int.\ J.\ Mod.\ Phys.\ {\bf A10}, 3497 (1995).
\bibitem{abott}
L.~F.~Abbott,
  Nucl.\ Phys.\ {\bf B185}, 189 (1981); Acta Phys.\ Polon.\ {\bf B13}, 33 (1982).
\bibitem{narison}
S.~Narison and R.~Tarrach, \plb{125}{217}{1983}.
\bibitem{morozov}
  A.~Y.~Morozov,
  Sov.\ J.\ Nucl.\ Phys.\  {\bf 40}, 505 (1984); Preprints ITEP-190, 191 (1983).
\bibitem{gammaj1}
 X.~-D.~Ji and M.~J.~Musolf,
  Phys.\ Lett.\ {\bf B257}, 409 (1991); 
D.~J.~Broadhurst and A.~G.~Grozin,
  Phys.\ Lett.\ {\bf B267}, 105 (1991).
\bibitem{broad}
D.~J.~Broadhurst and A.~G.~Grozin,
  Phys.\ Lett.\ {\bf  B274}, 421 (1992).
\bibitem{ElKhadra}
  S.~Narison,
  Camb.\ Monogr.\ Part.\ Phys.\ Nucl.\ Phys.\ Cosmol.\  {\bf 17}, 1 (2002).
\bibitem{Ball:1993xv} 
  P.~Ball and V.~M.~Braun,
Phys.\ Rev.\ D {\bf 49}, 2472 (1994)  [hep-ph/9307291].  
\bibitem{neubert2}
M.~Neubert, 
\prd{46}{1076}{1992}.
\bibitem{naka}
  K.~Nakamura {\it et al.}  [Particle Data Group],
  J.\ Phys.\ G {\bf 37}, 075021 (2010).
\bibitem{Politzer:1980me}
  H.~D.~Politzer,
  Nucl.\ Phys.\ {\bf B172}, 349 (1980);
 J.~C.~ Collins, {\it Renormalization}, (Cambridge
University Press,  1984);
J.~Kodaira and K.~Tanaka,
  Prog.\ Theor.\ Phys.\  {\bf 101}, 191 (1999).
\bibitem{kodaira}
J.~Kodaira, Y.~Yasui and T.~Uematsu,
  Phys.\ Lett.\ {\bf  B344}, 348 (1995);
J.~Kodaira, Y.~Yasui, K.~Tanaka and T.~Uematsu,
  Phys.\ Lett.\ {\bf  B387}, 855 (1996);
Y.~Koike and K.~Tanaka,
  Phys.\ Rev.\ {\bf  D51}, 6125 (1995);
Y.~Koike and N.~Nishiyama,
  Phys.\ Rev.\ {\bf D55}, 3068 (1997);
H.~Kawamura, T.~Uematsu, J.~Kodaira and Y.~Yasui,
  Mod.\ Phys.\ Lett.\ {\bf  A12}, 135 (1997);
H.~Kawamura,
  Z.\ Phys.\ {\bf C75}, 27 (1997).
\bibitem{Kodaira:1997ig}
E.~V.~Shuryak and A.~I.~Vainshtein,
  Nucl.\ Phys.\ {\bf B199}, 451 (1982); {\bf B201}, 141 (1982);
A.~P.~Bukhvostov, E.~A.~Kuraev and L.~N.~Lipatov,
Sov.\ Phys.\ JETP {\bf 60}, 22 (1984);
I.~I.~Balitsky and V.~M.~Braun,
  Nucl.\ Phys.\ {\bf B311}, 541 (1989);
  J.~Kodaira, T.~Nasuno, H.~Tochimura, K.~Tanaka and Y.~Yasui,
  Prog.\ Theor.\ Phys.\  {\bf 99}, 315 (1998).
\bibitem{Grozin:1996hk}
  A.~G.~Grozin and M.~Neubert,
  Nucl.\ Phys.\ {\bf B495}, 81 (1997).
\bibitem{BBL}
P.~Ball, V.~M.~Braun and A.~Lenz, 
JHEP\ {\bf 0605}, 004 (2006).
\end{thebibliography}
